\documentclass[11pt]{article}

\usepackage[margin=1in]{geometry}
\usepackage{authblk}
\usepackage{amsmath, amssymb}
\usepackage{graphicx}
\usepackage{hyperref}
\usepackage{titlesec}
\usepackage{setspace}
\usepackage[labelfont=bf]{caption}
\usepackage{subcaption} 
\usepackage{cite}
\usepackage{xcolor}
\usepackage{adjustbox}
\usepackage{subscript}
\usepackage{tablefootnote}
\usepackage{threeparttable}
\usepackage{booktabs}
\usepackage{quantikz}
\usepackage{tikz}
\usepackage[normalem]{ulem}

\titleformat{\section}{\large\bfseries}{\thesection}{1em}{}
\titleformat{\subsection}{\normalsize\bfseries}{\thesubsection}{1em}{}

\title{Accelerating Quantum Simulations of Materials Through Parameter and Ansatz Transfer Strategies}

\author[1]{Saurabh Shivpuje}
\author[2]{Vinit Singh}
\author[3]{Manas Sajjan}
\author[2,4]{Sabre Kais}

\affil[1]{James Tarpo Jr. and Margaret Tarpo
Department of Chemistry, Purdue University, West Lafayette, Indiana 47907, USA}
\affil[2]{Department of Electrical and Computer Engineering, North Carolina State University, Raleigh, NC 27606}
\affil[3]{National Center for Computational Sciences, Oak Ridge National Laboratory, Oak Ridge, Tennessee 37831, USA}
\affil[4]{Department of Chemistry, North Carolina State University, Raleigh, NC 27695, USA} 

\date{}

\begin{document}
\maketitle

\begin{abstract}

Quantum computing has emerged as a promising approach for electronic-structure calculations, yet practical condensed-matter applications often require solving large families of closely related Hamiltonians arising from $\mathbf{k}$-point sampling, compositional variations, defect configurations, and changes in system size. Here, we develop strategies for transferring variational information between related Hamiltonians to exploit these correlations and accelerate variational quantum simulations. For Hamiltonians of the same embedded size, we transfer optimized variational parameters between related calculations. For Hamiltonians of different sizes, we introduce Augmented Ansatz Reuse for Target Initialization (AARTI), which transfers compatible Pauli-generator structure and optimized source parameters from a previously solved Hamiltonian, then augments the transferred ansatz with target-specific generators to initialize optimization of a larger target Hamiltonian. Our workflow combines first-principles density functional theory, Wannier downfolding, and embedded tight-binding Hamiltonians with variational quantum algorithms and neural-network quantum states, with quantum-circuit simulations implemented using NVIDIA's CUDA-Q platform. Using Li$_x$CoO$_2$ as a representative battery cathode material, we construct and simulate Hamiltonian families spanning multiple delithiation levels, dense $\mathbf{k}$-point meshes, and increasing Hamiltonian sizes. The proposed parameter-transfer and ansatz-transfer approaches substantially reduce the number of optimization steps required to solve large families of related Hamiltonians while maintaining the common target accuracy, with lower terminal errors observed in several cases. The calculated evolution of the electronic structure with delithiation exhibits pronounced non-rigid-band behavior consistent with experimental observations. These results establish a framework for exploiting similarity among related Hamiltonians in quantum materials simulations and demonstrate how variational-information transfer can improve the efficiency and scalability of quantum workflows for realistic condensed-matter systems.

\end{abstract}

\section{Introduction}

Recent work in quantum computing has identified chemistry and materials science as promising application areas for electronic structure calculations~\cite{daley2022practical,rawat2025analyzing,google2025observation,zhang2025quantum,wu2024variational,nassir2025quantum,mazzola2024quantum,duriez2503computing,bauer2020quantum}. Many studies have proposed end-to-end frameworks that combine first-principles downfolding~\cite{alvertis2025compressing}, variational quantum eigensolvers~\cite{selisko2025dynamical,choudhary2021quantum}, and quantum machine learning~\cite{sajjan2021quantum}. The growing demand for improved materials continues to outpace experimental discovery, which remains constrained by high costs and time-consuming synthesis and characterization. As a result, the combinatorially large space of chemical substitutions and defect configurations is impractical to explore solely through experimental trial and testing~\cite{van2004first,jain2013commentary,lesar2013introduction,broberg2023high}. In condensed matter simulations, families of closely related Hamiltonians naturally arise from $\mathbf{k}$-point sampling, compositional variations, defect configurations, and changes in system size. Such Hamiltonian families provide an opportunity to reuse information between related quantum calculations. In this work, we develop strategies for transferring variational information between related Hamiltonians, including optimized parameters for fixed-size problems and Augmented Ansatz Reuse for Target Initialization (AARTI) for cross-size ansatz transfer, and demonstrate their application using Li$_x$CoO$_2$ (LCO), a prototypical lithium-ion battery cathode material.

LCO is a layered oxide material first reported in 1980 by Goodenough and co-workers as a rechargeable battery cathode~\cite{mizushima1980lixcoo2}, and it was later commercialized in lithium-ion batteries in 1991~\cite{manthiram2021layered}. Since then, LCO and related layered oxide materials such as LiNiO$_2$ and Li$_2$MnO$_3$ have attracted significant research attention~\cite{dahn1994thermal,papp2021comparison}. During lithium intercalation and deintercalation, the formation of Li vacancies alters the oxidation states of transition-metal atoms and can induce structural phase changes. These processes strongly influence key battery properties such as energy density, thermal stability, and cycle life~\cite{lin2024structural}. Despite substantial improvements over the past decades, continued interest in higher-performance battery materials motivates further research of these systems~\cite{radin2017narrowing}. Because LCO serves as a prototypical layered oxide cathode material, it provides a useful model system for studying such phenomena~\cite{fantin2023self,yang2021insights}.

In this study, we focus on the band structure of LCO across different lithium concentrations. Experimental and computational studies have reported a wide range of band-gap values for LCO~\cite{van1991electronic,ensling2010electronic,radha2021optical,ghosh2007structure,andriyevsky2014electronic}. However, most studies focus on the pristine material or the fully lithiated and fully delithiated end members, while intermediate intercalation states remain less explored~\cite{kang2019electrical,banifarsi2022optical}. Modeling LCO for intermediate lithium concentrations typically requires supercell calculations~\cite{chakraborty2018predicting,laubach2009changes,li2015halogen}. Exploring a wider range of lithium concentrations and defect configurations often necessitates large supercells to reduce artificial interactions between periodically repeated defects and to capture longer-range structural and chemical variations~\cite{wright2006comparison,freysoldt2022limitations,freysoldt2014first}. Furthermore, the number of possible defect arrangements grows rapidly, leading to a significant increase in computational cost. Each increase in supercell size multiplies the number of atoms, electrons, orbitals, and bands involved in the calculation. Methods that can efficiently transfer optimized variational information across related Hamiltonians may therefore substantially improve the practicality of quantum simulations for realistic condensed-matter systems. Qubit-based quantum computing may complement existing computational workflows as quantum hardware continues to improve.

Recent progress by Alvertis et al.~\cite{alvertis2025compressing} demonstrated how \textit{ab initio} downfolding can be used to construct effective Hamiltonians and compute band structures. In this work, we extend such workflows to enable band-gap calculations. Our approach incorporates Wannierization and projection onto selected orbitals to separate valence and conduction bands~\cite{marzari2012maximally}. Wannierization is widely used for constructing tight-binding Hamiltonians. Using Fourier transforms, these Hamiltonians can interpolate band structures onto denser $\mathbf{k}$-point meshes than those used in the original density functional theory calculations~\cite{marzari1997maximally}. For electronic-structure properties evaluated on dense $\mathbf{k}$ meshes, the resulting number of related Hamiltonians can become large, motivating strategies that reduce repeated optimization across the mesh~\cite{lu2020ab}.

Most previous quantum computing studies in materials science employ the variational quantum eigensolver (VQE) as the primary algorithm~\cite{peruzzo2014variational,selisko2025dynamical,choudhary2021quantum,skogh2023accelerating}. In addition to VQE, we explore a neural-network quantum-state approach that has not yet been widely applied to these problems. Specifically, we consider a neural quantum state (NQS) represented by a restricted Boltzmann machine (RBM) and optimized using variational Monte Carlo (VMC) with Markov chain Monte Carlo (MCMC) sampling incorporating parameterized quantum circuits~\cite{sajjan2026polynomially}. We denote this workflow NQS/VMC/Adam. Because condensed-matter simulations often involve large families of related Hamiltonians spanning $\mathbf{k}$ points, compositions, defect configurations, and embedded system sizes, transferring variational information from previously solved source Hamiltonians to related target Hamiltonians provides a natural route to reduce repeated optimization effort~\cite{sajjan2021quantum,skogh2023accelerating,villar2026transfer}. Related cross-size transfer concepts have been explored in the Quantum Approximate Optimization Algorithm (QAOA)~\cite{galda2023similarity}, where optimized parameters from smaller problem instances can be reused for larger instances, and in quantum learning models through sub-network initialization~\cite{kubal2025quantum} as the number of qubits increases. In contrast, AARTI provides a Hamiltonian-informed strategy for cross-size transfer: compatible Hamiltonian-aligned Pauli generators are identified and embedded into the target representation, their optimized parameters are inherited, and the transferred ansatz is augmented with target-specific generators. We apply fixed-size parameter transfer and AARTI to Wannier-derived Li$_x$CoO$_2$ Hamiltonians across $\mathbf{k}$ points, lithium compositions, and increasing Hamiltonian sizes, and evaluate their effectiveness in reducing repeated variational optimization while retaining the target accuracy used throughout this work. We also analyze the evolution of the electronic structure with delithiation and its non-rigid-band behavior.

\section{Methods}
The overall computational workflow employed in this study consists of four stages: generation of defect-containing Li$_x$CoO$_2$ structures using first-principles calculations, construction of Wannier-based embedded Hamiltonians, solution of these Hamiltonians using quantum algorithms, and acceleration of calculations through variational parameter transfer. The key components of this workflow are described below.
\subsection{Density Functional Theory (DFT) Supercell Construction}

DFT calculations were carried out using the Quantum ESPRESSO package~\cite{qe}. The electronic wavefunctions were expanded in a plane-wave basis using projector augmented-wave (PAW) datasets, with kinetic-energy cutoffs of 60 Ry for the wavefunctions and 445 Ry for the charge density~\cite{basis_set}. Electronic occupations were treated using a smearing scheme with a width of 0.02 Ry. A Hubbard $U$ parameter of 3 eV was applied to the Co $3d$ states~\cite{chakraborty2018predicting}.

A $2 \times 1 \times 1$ supercell of LiCoO$_2$ containing 24 atoms was used for the majority of the calculations reported in this work. To investigate transfer across increasing Hamiltonian sizes, additional supercells up to $4 \times 4 \times 1$ were also considered. Different lithium concentrations were generated by systematically removing lithium atoms to construct Li$_x$CoO$_2$ with compositions $x = 0.00$, $0.17$, $0.33$, $0.50$, $0.67$, $0.83$, and $1.00$. For each composition, all symmetrically distinct lithium-vacancy configurations within the supercell were considered, and the lowest-energy relaxed structure was selected for subsequent Wannierization and quantum simulations. Brillouin-zone integrations were performed using Monkhorst--Pack
$\mathbf{k}$-point meshes appropriate for the supercell size and calculation
objective. The majority of the calculations employed a
$4 \times 8 \times 2$ $\mathbf{k}$-point grid, while additional meshes were
used for selected larger-supercell calculations and electronic-structure
analyses.

\begin{figure}[t]
    \centering
    \includegraphics[width=0.50\linewidth]{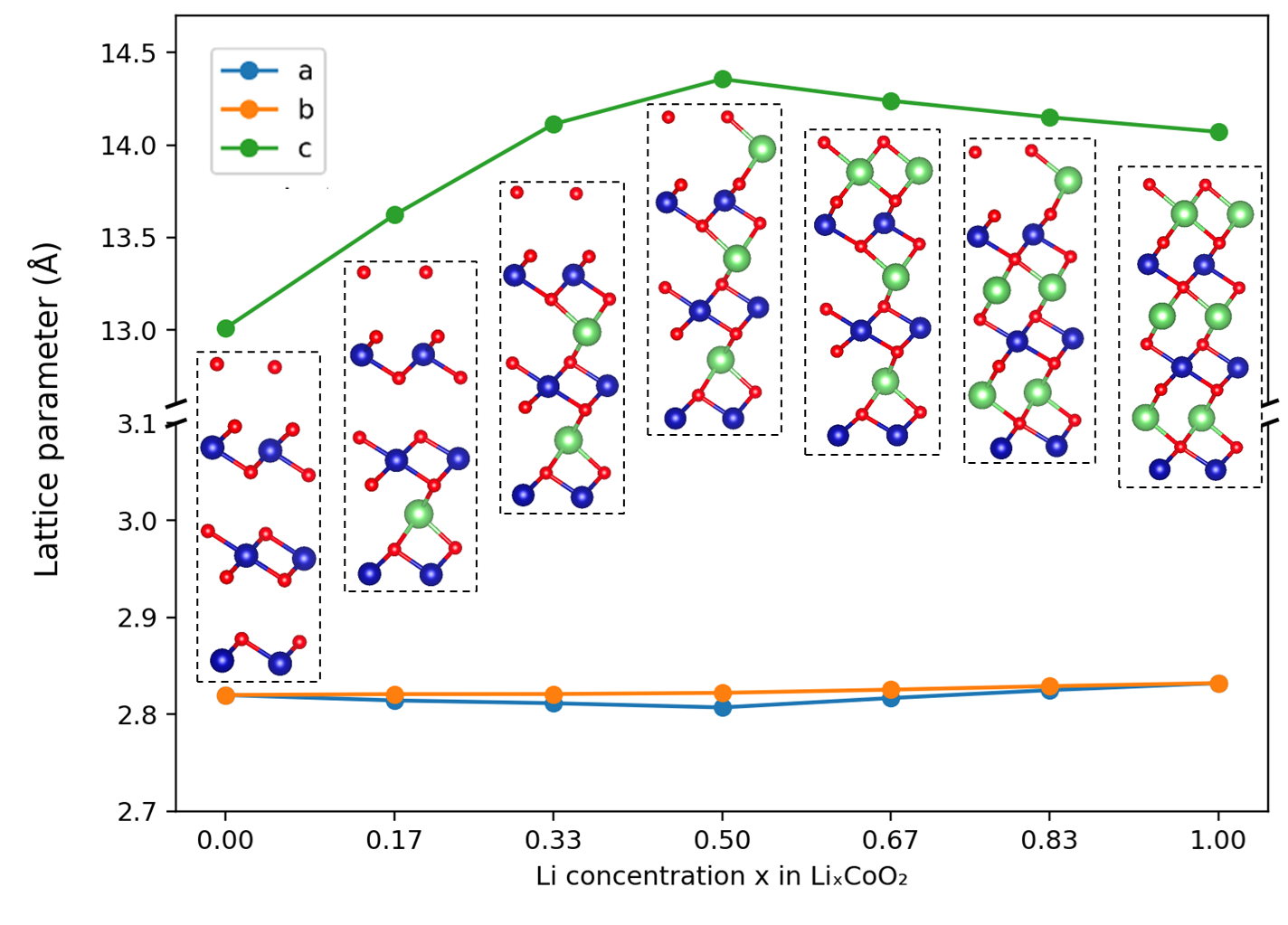}
    \caption{\textbf{Structural evolution of Li$_x$CoO$_2$ with delithiation.}
    Evolution of the lattice parameters $a$, $b$, and $c$ as a function of lithium concentration $x$. Representative relaxed supercell structures corresponding to the lowest-energy lithium-vacancy configurations are shown as insets for each composition. The selected structures span the delithiation pathway from CoO$_2$ ($x=0.00$) to fully lithiated LiCoO$_2$ ($x=1.00$), providing a family of defect-containing configurations with varying local chemical environments that serve as the starting point for the Wannier downfolding and quantum simulations performed in this work.}
    \label{fig:fig1a}
\end{figure}

The resulting set of relaxed structures spans the delithiation pathway from
fully lithiated LiCoO$_2$ to CoO$_2$, thereby providing a family of
defect-containing configurations with varying local chemical environments.
Representative relaxed structures and the corresponding evolution of the lattice
parameters are shown in Fig.~\ref{fig:fig1a}.

\subsection{Wannier Downfolding and Embedded Hamiltonians}

The electronic structure of a periodic crystal is naturally described in terms of Bloch states. To obtain a compact representation suitable for quantum simulation, maximally localized Wannier functions were constructed from the DFT electronic structure using \texttt{Wannier90}~\cite{pizzi2020wannier90}. In this work, we focused on the bands near the Fermi level that are primarily associated with the Co $3d$ states.

The Wannier Hamiltonian was expressed in terms of real-space hopping matrix elements between localized orbitals, enabling interpolation of the electronic structure on arbitrary $\mathbf{k}$ meshes through Fourier transformation. This representation provides an efficient route for generating large families of related Hamiltonians without requiring repeated first-principles calculations.

Using the projection scheme implemented in \texttt{Wannier90}, the Co $3d$ manifold was separated into the lower-energy $t_{2g}$ and higher-energy $e_g$ subspaces. Independent embedded Hamiltonians corresponding to these manifolds were then constructed and evaluated along both high-symmetry $\mathbf{k}$ paths and dense $10 \times 10 \times 10$ $\mathbf{k}$ meshes. The resulting Hamiltonian families span multiple lithium concentrations and $\mathbf{k}$ points, providing the foundation for the transfer strategies explored in this work.

\subsection{Quantum Solvers}

For consistency, the four primary variational workflows used in the transfer benchmarks are denoted according to their state representation or ansatz, variational framework, and optimizer: HEA/VQE/COBYLA, HEA/VQE/Adam, HPA/VQE/Adam, and NQS/VMC/Adam. Here, HEA denotes the hardware-efficient ansatz, HPA the Hamiltonian-aligned Pauli ansatz, and COBYLA denotes Constrained Optimization BY Linear Approximations. The NQS/VMC/Adam calculations
employ the quantum-enabled MCMC sampling procedure described below. The quantum simulations were implemented using NVIDIA's CUDA-Q, an
open-source, qubit-agnostic platform for hybrid quantum--classical computing
with graphics processing unit (GPU)-accelerated simulation capabilities. CUDA-Q was used to execute the
quantum-circuit components of the variational workflows considered here.

To obtain the eigenvalues of the embedded Hamiltonians, we employed two classes of variational approaches: the Variational Quantum Eigensolver (VQE) and the NQS/VMC/Adam neural-network quantum-state approach described below.
 For the lowest $e_g$ conduction-band edge, the variational algorithms minimized
$\hat{H}(\mathbf{k})$. For the highest $t_{2g}$ valence-band edge, they minimized
$-\hat{H}(\mathbf{k})$; the reported valence energy was then obtained as
$-E_{\min}[-\hat{H}(\mathbf{k})]$. Exact diagonalization references were evaluated
using the corresponding extremal eigenvalues for benchmarking accuracy and assessing convergence of the variational calculations. For the primary \(2\times1\times1\) supercell calculations, the \(e_g\) and \(t_{2g}\) embedded Hamiltonian executions were presented using four and five qubits, respectively. The variational circuit prepares an \(N_q\)-qubit statevector \( |\psi(\boldsymbol{\theta})\rangle \) and the number of qubits \(N_q\) is therefore determined by the dimension of the embedded Hamiltonian representation.

\subsubsection{Variational Quantum Eigensolver}

In VQE~\cite{peruzzo2014variational}, a parameterized quantum circuit prepares a trial state,
\begin{equation}
|\psi(\boldsymbol{\theta})\rangle =
U(\boldsymbol{\theta}) |0\rangle,
\end{equation}
where $U(\boldsymbol{\theta})$ is a parameterized unitary operator and $\boldsymbol{\theta}$ denotes the set of variational parameters. The objective function is given by
\begin{equation}
E(\boldsymbol{\theta}) =
\langle \psi(\boldsymbol{\theta}) |
\hat{H}
| \psi(\boldsymbol{\theta}) \rangle,
\end{equation}
which is minimized using a classical optimizer.

Two ansatz constructions were considered. First, hardware-efficient ansatz (HEAs) composed of repeated layers of parameterized single-qubit rotations and nearest-neighbor entangling gates were employed to investigate transfer behavior across different optimizers. The variational parameters were optimized using both the gradient-free COBYLA optimizer and the Adam optimizer. For Adam-based calculations, gradients were evaluated using the parameter-shift rule.

Second, we employed a Hamiltonian-aligned Pauli ansatz constructed from the Pauli decomposition of the embedded Wannier Hamiltonians. For transfer studies involving Hamiltonians of the same embedded size, the operator structure derived from the first $\Gamma$-point Hamiltonian of fully lithiated LiCoO$_2$ ($x=1.00$) was retained, while the associated variational parameters were transferred and reoptimized for related Hamiltonians. The extension of this approach to cross-size transfer through AARTI is described below. The VQE circuits and associated quantum simulations were executed using
CUDA-Q. Additional implementation details, including circuit specifications
and ansatz constructions, are provided in the Supporting Information.

\subsubsection{Neural Quantum State with Variational Monte Carlo
}

As an alternative variational solver, we employed an RBM-based neural quantum state optimized using variational Monte Carlo (VMC), with configurations generated through Markov chain Monte Carlo (MCMC) sampling incorporating parameterized quantum circuits~\cite{sajjan2026polynomially}. In this framework, the many-body wavefunction is represented as
\begin{equation}
\psi(\mathbf{s}) =
\exp\left(
\sum_i a_i s_i
\right)
\prod_j
\cosh\left(
b_j + \sum_i W_{ij} s_i
\right),
\end{equation}
where $\mathbf{s}$ denotes a computational basis configuration, and $a_i$, $b_j$, and $W_{ij}$ are variational parameters corresponding to visible biases, hidden biases, and coupling weights.

The energy expectation value is evaluated using variational Monte Carlo sampling, with configurations generated according to the probability distribution $|\psi(\mathbf{s})|^2$. Parameterized quantum circuits were used to enhance the sampling procedure, enabling estimation of observables and stochastic optimization of the network parameters. Additional details of the sampling circuits, hyperparameter settings, and optimization procedures are provided in the Supporting Information.

To ensure a consistent convergence criterion across all simulations, optimization was terminated when the variational energy reached within 0.03 eV of the corresponding exact eigenvalue and remained within this threshold for five consecutive iterations. The exact eigenvalue is used here only to define a uniform benchmarking criterion across methods and is not intended as a convergence oracle for practical quantum calculations. Throughout this work, acceleration is quantified by the number of optimization
steps required to satisfy the common convergence criterion. Accordingly, the
reported speedups refer to convergence-step reductions rather than wall-time
or total computational-resource speedups, since the cost of an individual
optimization step differs among the variational workflows. The
convergence-step speedup is defined as the ratio of the mean number of
optimization steps required without transfer to that required with transfer,
evaluated over the same set of target Hamiltonians. Targets that do not satisfy
the convergence criterion within the allotted optimization budget are assigned
the corresponding maximum number of optimization steps.

\subsection{Transfer Strategies}\label{sec:transfer-strategies}

The dense $\mathbf{k}$-point meshes and multiple lithium concentrations considered in this work give rise to large families of closely related embedded Hamiltonians. Solving each Hamiltonian independently using variational approaches can require substantial optimization effort, motivating the use of transfer strategies that exploit correlations between previously solved and target problems.

For Hamiltonians of the same embedded size, we investigated transfer between neighboring \(\mathbf{k}\) points within a given composition and between corresponding \(\mathbf{k}\) points across different lithium concentrations. These fixed-size transfer calculations were performed using the \(2\times1\times1\) supercell, corresponding to four-qubit \(e_g\) and five-qubit \(t_{2g}\) Hamiltonians. Thus, parameter transfer in these calculations occurs between source and target Hamiltonians represented using the same number of qubits. Two transfer protocols were considered. In serial transfer, optimized parameters from one calculation were propagated sequentially to initialize subsequent calculations. In parallel transfer, parameters obtained from a reference calculation were distributed simultaneously to multiple target Hamiltonians. These strategies were examined across the VQE-based workflows and NQS/VMC/Adam to assess the extent to which transfer behavior depends on the underlying solver, ansatz construction, and optimization procedure.

Transfer between different lithium concentrations was implemented by mapping optimized parameters from one composition to the corresponding $\mathbf{k}$ points of neighboring compositions. The resulting transfer trends provide insight into whether similarities in local chemical environments can be leveraged to accelerate convergence across defect-containing Hamiltonian families.

The transfer approaches described above operate within fixed embedded Hamiltonian sizes. To extend transfer beyond this setting, we further considered AARTI, in which operator-level information derived from previously solved Hamiltonians is reused to initialize variational calculations for larger embedded systems. The AARTI methodology and its application to cross-size transfer are described below.

\subsubsection{Augmented Ansatz Reuse for Target Initialization (AARTI)}

\begin{figure}[ht]
    \centering
    \includegraphics[width=1\linewidth]{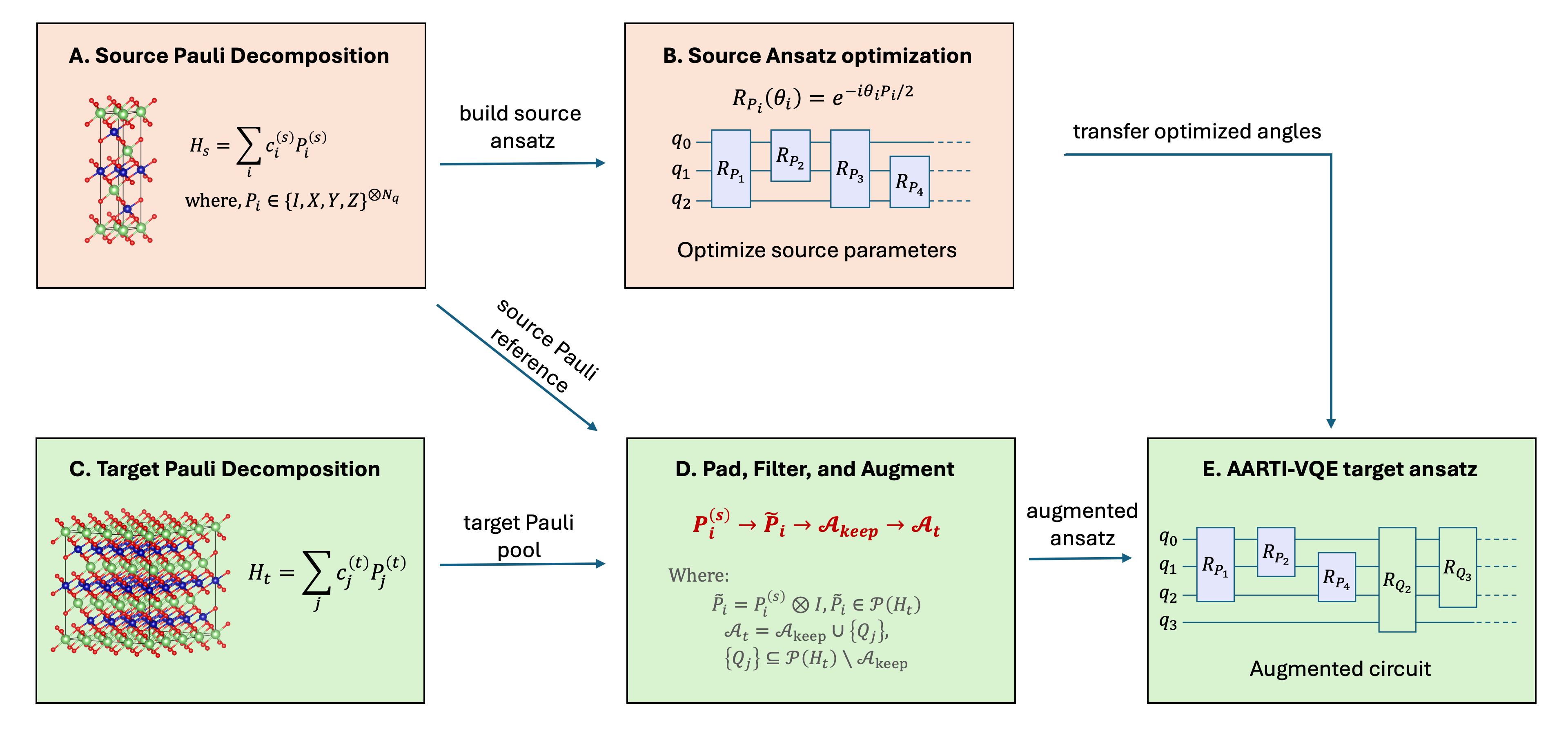}
    \caption{\textbf{Augmented Ansatz Reuse for Target Initialization (AARTI) workflow.} (A) The source Hamiltonian is decomposed into Pauli operators to construct a Hamiltonian-aligned source ansatz. (B) The source ansatz is optimized to obtain variational parameters. (C) The target Hamiltonian is decomposed into its Pauli operator pool. (D) Source Pauli generators are padded, matched to the target Hamiltonian, and augmented with target-specific operators to construct the augmented target ansatz. (E) Optimized parameters associated with the reused operators are transferred to initialize the target ansatz, while newly introduced operators are assigned independent parameters and subsequently optimized.}
    \label{fig:aarti_vqe}
\end{figure}

To enable transfer between embedded Hamiltonians of different sizes, we employed AARTI. Unlike the transfer protocols described above, which reuse variational parameters within a fixed Hamiltonian dimension, AARTI transfers both operator-level information and optimized parameters across changing embedded representations.

For each embedded Hamiltonian, the Hamiltonian was first expressed as a Pauli expansion,
\begin{equation}
\hat{H}=\sum_j c_j \hat{P}_j,
\end{equation}
where $\hat{P}_j$ denotes a tensor product of single-qubit Pauli operators and $c_j$ is the corresponding coefficient. The identity term was excluded, and a subset of non-identity Pauli strings was selected to define an ordered ansatz-generator set,
\begin{equation}
\mathcal{G}=
\left\{
\hat{P}_1,\hat{P}_2,\ldots,\hat{P}_{N_P}
\right\}.
\end{equation}

The corresponding trial state was prepared as a sequence of Pauli rotations,
\begin{equation}
|\psi(\boldsymbol{\theta})\rangle=
\prod_{j=1}^{N_P}
\exp\left(
-\frac{i}{2}\theta_j \hat{P}_j
\right)
|0\rangle^{\otimes N_q},
\end{equation}
where \(N_q\) denotes the number of qubits in the embedded Hamiltonian representation and each Pauli generator contributes a single variational parameter.

For cross-size transfer, Pauli generators from a previously optimized source ansatz were embedded into the target qubit register through identity padding and compatibility matching. The retained generators formed the transferred generator set, which was subsequently augmented with target-specific Pauli generators. Optimized source parameters were used to initialize the corresponding transferred parameters, while parameters associated with newly introduced generators were initialized according to the target-Hamiltonian construction procedure. The resulting augmented target ansatz was then optimized for the larger target Hamiltonian. Additional details regarding generator selection, padding strategies, transfer variants, and implementation settings are provided in the Supporting Information.

\section{Results and Discussion}

Before evaluating the transfer strategies, we verified that the embedded Hamiltonians reproduce the expected electronic-structure behavior of Li$_x$CoO$_2$. In particular, the exact solutions reproduce the experimentally observed non-rigid evolution with delithiation~\cite{ensling2014nonrigid,tong2025first,xiong2012atomic}, while the direct and indirect band gaps of 2.24 and 2.19~eV obtained for fully lithiated LiCoO$_2$ are also in agreement with reported experimental values~\cite{kushida2001optical,liu2015electronic,ghosh2007structure,balakrishnan2019studies,rao2009optical,van1991electronic}. The corresponding composition-dependent band-edge analysis is provided in the Supporting Information. We therefore focus the discussion below on the efficiency of parameter and ansatz transfer across these physically relevant Hamiltonian families.

\subsection{Parameter Transfer Along the k-Path}

We first examined parameter transfer between neighboring Hamiltonians along a
high-symmetry $\mathbf{k}$ path within a fixed composition, fully lithiated
LiCoO$_2$ ($x=1.00$). These calculations use the $2\times1\times1$
supercell, corresponding to four qubits for the $e_g$ manifold and five
qubits for the $t_{2g}$ manifold. Because the Hamiltonian dimension is
unchanged along the $\mathbf{k}$ path, optimized parameters are transferred
between systems having the same number of qubits. The
$\Gamma$--$M$--$K$--$\Gamma$ path contains 91 $\mathbf{k}$ points, indexed
as $k_{000},\ldots,k_{090}$, with $k_{000}$ corresponding to the initial
$\Gamma$ point. Each point defines a distinct embedded Hamiltonian for the
$e_g$ and $t_{2g}$ manifolds. The $k_{000}$ calculation was cold-started
for each workflow, and the serial and parallel transfer protocols defined in
Sec.~\ref{sec:transfer-strategies} were compared over the remaining 90
$\mathbf{k}$ points against independent no-transfer calculations.

\begin{figure}[ht]
    \centering
    \includegraphics[width=\linewidth]
    {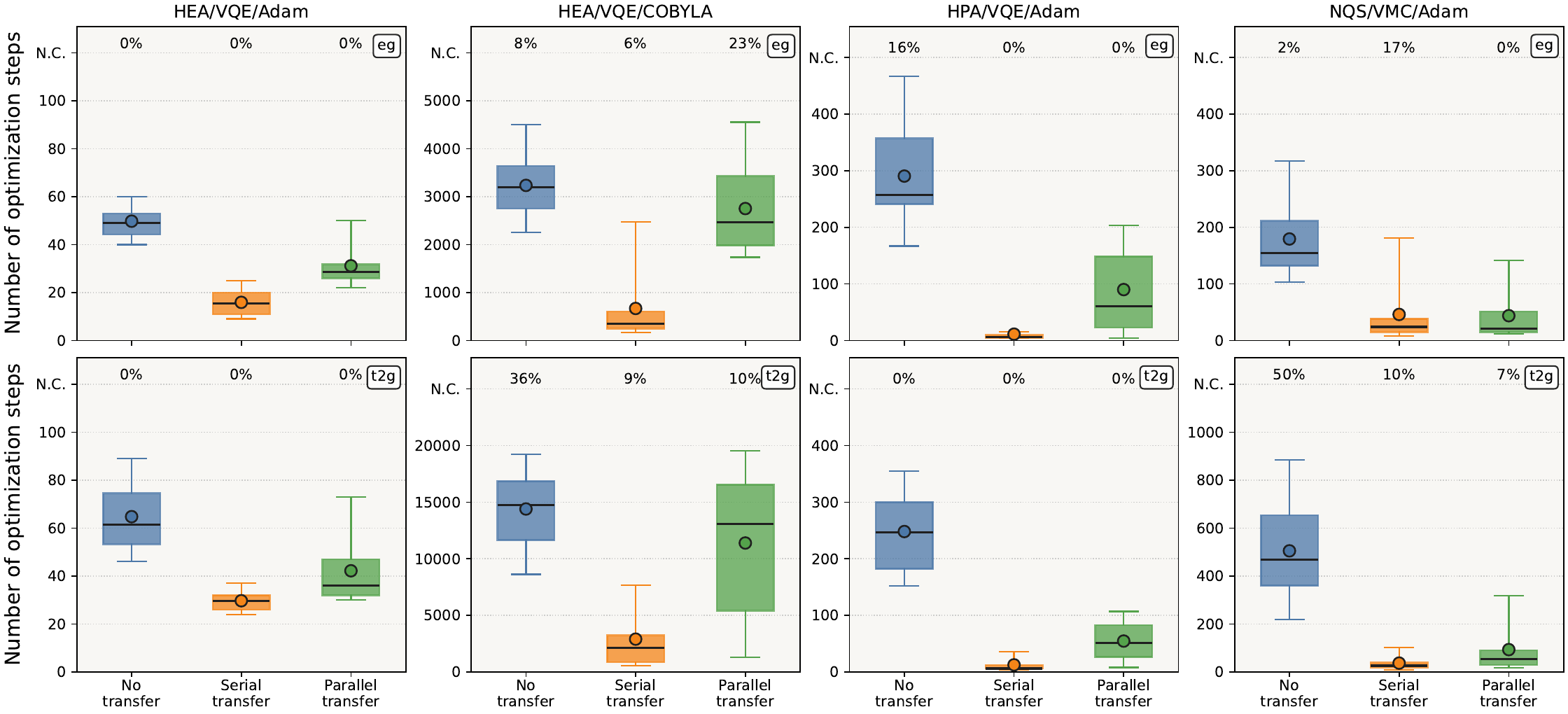}
    \caption{
    \textbf{Parameter transfer along the high-symmetry $\mathbf{k}$ path.}
    Optimization steps required for no-transfer, serial-transfer, and
    parallel-transfer initialization over $k_{001}$--$k_{090}$ along the
    high-symmetry $\mathbf{k}$ path. Results are shown for
    HEA/VQE/COBYLA, HEA/VQE/Adam, HPA/VQE/Adam, and NQS/VMC/Adam.
    The upper and lower rows correspond to the $e_g$ and $t_{2g}$
    manifolds, respectively. Open circles denote the mean, and percentages
    above the box plots indicate the fraction of calculations that did not
    satisfy the convergence criterion. The cold-start $k_{000}$ calculation
    is not included in the distributions.
    }
    \label{fig:kpoint-transfer}
\end{figure}

Figure~\ref{fig:kpoint-transfer} reveals a clear dependence of transfer
behavior on the variational workflow. For all three VQE workflows, serial
transfer provides the largest and most consistent reduction in optimization
steps for both orbital manifolds. The trend persists across changes in both
optimizer and ansatz, indicating that sequential reuse between neighboring
$\mathbf{k}$-point Hamiltonians is particularly effective for the VQE
calculations considered here. HPA/VQE/Adam exhibits the strongest response,
with the mean optimization cost decreasing from about 291 to 11 steps for
$e_g$ and from 248 to 12 steps for $t_{2g}$ under serial transfer. Parallel
transfer is also beneficial in several cases, but is generally less efficient
than serial transfer.

The comparison between the two HEA/VQE workflows further shows that the
optimizer strongly influences both the absolute optimization cost and the
effectiveness of parameter reuse. For HEA/VQE/COBYLA, serial transfer reduces
the mean optimization cost from about 3235 to 669 steps for $e_g$ and from
14393 to 2896 steps for $t_{2g}$. In contrast, parallel transfer is
considerably less robust for the $e_g$ manifold, where approximately 23\%
of the targets remain unconverged. These comparisons illustrate how the
observed transfer trends can be used to identify an appropriate
ansatz--optimizer--transfer combination before extending the calculation to
larger Hamiltonian families. Detailed numerical statistics corresponding to
Fig.~\ref{fig:kpoint-transfer} are provided in the Supporting Information.

NQS/VMC/Adam displays a distinct transfer pattern. For the $e_g$ manifold,
parallel and serial transfer require similar optimization effort, with mean
costs of about 44 and 46 steps, respectively, while parallel transfer
eliminates the non-converged cases compared with approximately 17\% under
serial transfer. For $t_{2g}$, serial transfer is substantially more
efficient, requiring about 37 steps compared with 93 steps for parallel
transfer, whereas parallel transfer gives a slightly lower non-convergence
rate (about 7\% compared with 10\%). Thus, the preferred initialization
strategy depends not only on the variational framework but also on the
orbital manifold being considered.

The transfer trends in Fig.~\ref{fig:kpoint-transfer} were then used to
select a workflow-specific transfer strategy for completing the full
$\mathbf{k}$ path. HEA/VQE/Adam and HPA/VQE/Adam used serial transfer
without additional treatment. For HEA/VQE/COBYLA, serial transfer was used
as the primary strategy, and remaining unconverged points were reinitialized
using optimized parameters from a previously converged neighboring
$\mathbf{k}$ point. For NQS/VMC/Adam, parallel transfer was used as the
common primary strategy across the two manifolds because it eliminated the
unconverged cases for $e_g$ and produced the lower non-convergence fraction
for $t_{2g}$, although serial transfer required fewer $t_{2g}$ optimization
steps on average. Remaining unconverged $t_{2g}$ points were subsequently
initialized from successfully converged neighboring $\mathbf{k}$ points.

The performance of these selected workflows was evaluated over all 90
transferred targets, $k_{001}$--$k_{090}$. Non-converged runs were assigned
their full optimization-step limit, and when rescue was required, the total
workflow cost included both the failed primary attempt and the subsequent
rescue calculation. The convergence-step speedup was calculated as the ratio
of the mean cold-start cost to the mean selected-workflow cost over the same
90 targets. The resulting workflow-level convergence-step speedups are
summarized in Table~\ref{tab:speedups}.

\begin{table}[h]
\centering
\caption{\textbf{Solver specific transfer workflows and workflow level speedups.} Speedups are reported relative to independent
cold-start optimizations over the 90 transferred targets
$k_{001}$--$k_{090}$ of the 91-point
$\Gamma$--$M$--$K$--$\Gamma$ path. The initial $k_{000}$ calculation is
excluded because it serves as the transfer source. Non-converged runs are
assigned their full optimization-step limit, and rescued targets include
both the failed primary-run cost and the subsequent rescue-run cost.}
\label{tab:speedups}
\begin{tabular}{lccc}
\toprule
Method & Transfer strategy & $e_g$ speedup & $t_{2g}$ speedup \\
\midrule
NQS/VMC/Adam
& Parallel + neighboring-$\mathbf{k}$ rescue
& $4.25\times$ & $4.85\times$ \\
HEA/VQE/COBYLA
& Serial + backward-neighbor rescue
& $3.65\times$ & $3.67\times$ \\
HEA/VQE/Adam
& Serial neighboring-$\mathbf{k}$ transfer
& $3.13\times$ & $2.18\times$ \\
HPA/VQE/Adam
& Serial neighboring-$\mathbf{k}$ transfer
& $28.74\times$ & $19.90\times$ \\
\bottomrule
\end{tabular}
\end{table}

All selected workflows ultimately converged across the 90 transferred target
Hamiltonians. HPA/VQE/Adam exhibited the largest convergence-step reduction,
with speedups of $28.74\times$ and $19.90\times$ for the $e_g$ and $t_{2g}$
manifolds, respectively, while the other solver-specific workflows also
reduced the optimization effort relative to independent cold-start
calculations.

\begin{figure}[ht!]
    \centering
    \includegraphics[width=0.75\linewidth]
    {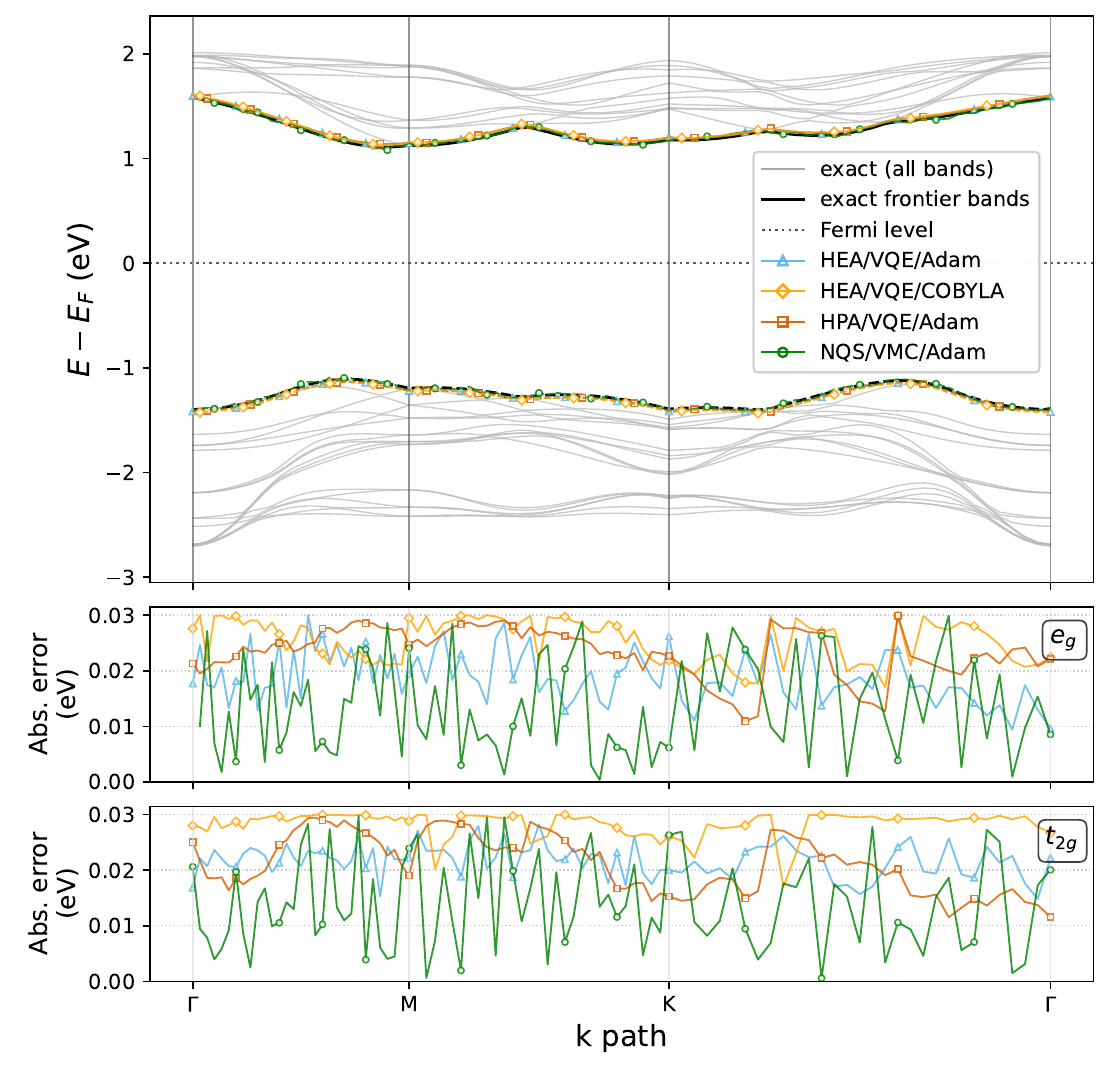}
    \caption{
    \textbf{Band-edge energies along the high-symmetry $\mathbf{k}$ path.}
    Energies obtained along the 91-point $\Gamma$--$M$--$K$--$\Gamma$ path
    for fully lithiated LiCoO$_2$ ($x=1.00$) using the solver-specific
    transfer workflows. Results from HEA/VQE/COBYLA, HEA/VQE/Adam,
    HPA/VQE/Adam, and NQS/VMC/Adam are shown together with the exact
    embedded-Hamiltonian band structure. All exact bands are shown as thin
    gray lines, while the exact frontier bands ($e_g$, $t_{2g}$) are
    emphasized in black. The colored variational curves include all
    $\mathbf{k}$ points; for visual clarity, hollow markers are displayed
    every sixth point with method-dependent starting offsets. Rescue is
    applied to unconverged HEA/VQE/COBYLA and NQS/VMC/Adam targets as
    described in the text. The lower two panels report the corresponding
    absolute deviations for the $e_g$ and $t_{2g}$ band edges separately.
    }
    \label{fig:kpath-bands}
\end{figure}

The resulting band-edge trajectories are shown in
Fig.~\ref{fig:kpath-bands}. The non-convergence fractions in
Fig.~\ref{fig:kpoint-transfer} reflect the initial transfer strategies,
whereas Fig.~\ref{fig:kpath-bands} shows the completed $\mathbf{k}$ path
after the required rescue calculations.

We also examined whether parameter transfer affected the terminal solution
quality under the common stopping criterion. Transfer maintained the common
target accuracy and, in several cases, produced lower terminal errors than
independent cold-start calculations. The improvement was most pronounced
for NQS/VMC/Adam, particularly for the $t_{2g}$ manifold, where transfer
reduced the large-error tail of the distribution. Because these errors are
evaluated at the stopping point rather than after exhaustive optimization,
they reflect the practical solution quality obtained within the adopted
workflow. Terminal-error distributions for all methods are provided in the
Supporting Information. Thus, parameter reuse provides a dual practical
benefit in these calculations: it reduces the optimization effort required
to reach the convergence criterion while maintaining the common target
accuracy, with lower terminal errors observed in several cases.
\subsection{Parameter Matched Comparison of Ansatz-Dependent Transfer}

The preceding comparison evaluates complete variational workflows and therefore
does not by itself isolate the origin of the particularly strong transfer response
observed for HPA/VQE/Adam. In particular, the workflows differ in ansatz
construction and in the number of variational parameters. To examine whether the
HPA transfer advantage can be explained primarily by its smaller parameter count,
we performed an additional parameter-matched comparison between
HEA/VQE/Adam, HPA/VQE/Adam, and a low-Hamiltonian-weight Pauli ansatz
(LPA/VQE/Adam). For this control, all three ansatzes contained 80 variational
parameters for the $e_g$ manifold and 150 parameters for the $t_{2g}$ manifold.
HPA and LPA were further constrained to have identical distributions of Pauli
generator support, while differing in the Hamiltonian coefficients used to select
the generators. HPA preferentially retained generators associated with
large-magnitude Hamiltonian coefficients, whereas LPA retained generators
associated with small-magnitude coefficients under the same parameter and
support constraints. Additional construction details are provided in the
Supporting Information.

\begin{figure}[ht]
    \centering
    \includegraphics[width=0.95\linewidth]
    {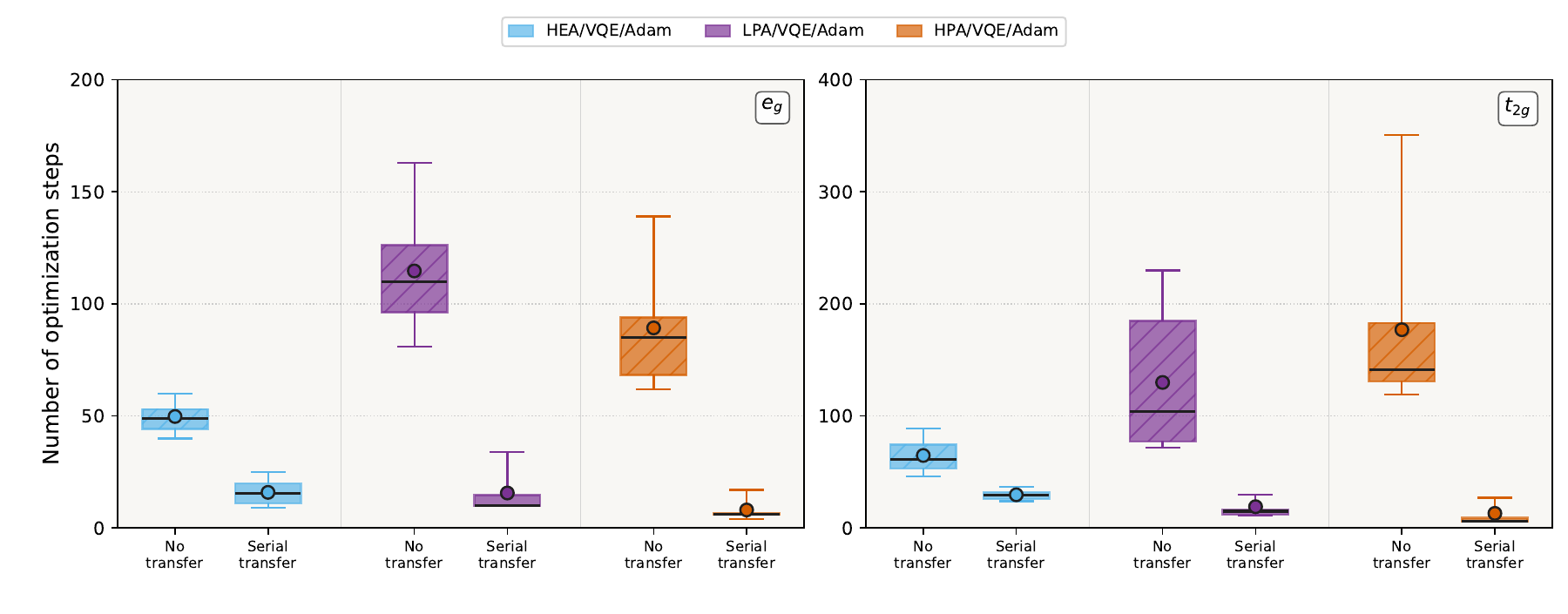}
    \caption{
    \textbf{Parameter-matched comparison of ansatz-dependent transfer behavior.}
    Optimization steps required for HEA/VQE/Adam, LPA/VQE/Adam, and
    HPA/VQE/Adam under no-transfer and serial-transfer initialization along
    $k_{001}$--$k_{090}$ of the high-symmetry $\mathbf{k}$ path.
    The ansatzes contain the same number of trainable variational parameters
    within each manifold: 80 for $e_g$ and 150 for $t_{2g}$.
    HPA and LPA additionally have identical distributions of one-, two-,
    three-, and four-qubit Pauli-generator support, as applicable to each
    manifold, while differing in the Hamiltonian-coefficient ranking used for
    generator selection. Open circles denote the mean, boxes show the
    interquartile range and median, and whiskers indicate the 5th--95th
    percentiles. The independently initialized $k_{000}$ calculation is not
    included in the distributions.
    }
    \label{fig:param-matched-transfer}
\end{figure}

Figure~\ref{fig:param-matched-transfer} shows that the strong transfer response of
HPA cannot be attributed solely to a reduced number of variational parameters.
Although the no-transfer HPA calculations are not uniformly less costly than the
parameter-matched HEA baseline, serial transfer reduces the HPA optimization
cost below that of HEA for both orbital manifolds. The distinction is therefore
not simply that HPA provides an easier cold-start optimization; rather, its
parameterization appears particularly amenable to reuse between neighboring
$\mathbf{k}$-point Hamiltonians.

LPA also exhibits a pronounced reduction in optimization cost upon switching
from no-transfer to serial-transfer initialization. Because HPA and LPA have the
same number of parameters and the same Pauli-support distribution, this common
response indicates that a substantial component of the transfer benefit is not
specific to selecting only the largest-magnitude Hamiltonian terms. Instead, the
result is consistent with a role for the shared multi-qubit Pauli-rotation
parameterization in making optimized parameters transferable between neighboring
Hamiltonians. HPA nevertheless retains the lowest serial-transfer optimization
cost, which is consistent with an additional contribution from
Hamiltonian-informed generator selection. A systematic separation of Pauli support, generator
ordering, and Hamiltonian-coefficient ranking is left for future investigation.

Taken together, these observations suggest that ansatzes built from shared
multi-qubit Pauli rotations may possess a parameterization that is
particularly amenable to transfer between related Hamiltonians. While the
present comparison does not isolate multi-qubit structure as the sole origin
of the improvement, it motivates further investigation of whether
multi-generator or multi-qubit ansatz constructions provide systematically
greater transferability than more local parameterizations.

Overall, these results demonstrate that similarities across related
$\mathbf{k}$-point Hamiltonians can substantially reduce repeated
variational optimization, while the most effective transfer strategy
depends on the ansatz, optimizer, variational framework, and orbital
manifold. We next examine whether parameter reuse remains effective when
the Hamiltonians differ not only in $\mathbf{k}$ point but also in
Li$_x$CoO$_2$ composition.

\subsection{Inter-Composition Parameter Transfer}

Having established the effectiveness of parameter transfer between neighboring
$\mathbf{k}$-point Hamiltonians within a fixed composition, we next examined
whether variational information could also be transferred across different
Li$_x$CoO$_2$ compositions. In this setting, optimized parameters obtained for
a Hamiltonian at one lithium concentration were used to initialize the
corresponding $\mathbf{k}$-point Hamiltonian at a neighboring lithium
concentration. We refer to this procedure as \emph{inter-composition parameter
transfer}. The transfer was performed sequentially along the delithiation
direction,
$x=1.00 \rightarrow 0.83 \rightarrow 0.67 \rightarrow 0.50
\rightarrow 0.33 \rightarrow 0.17 \rightarrow 0.00$, with optimized
parameters from each composition used to initialize the corresponding
$\mathbf{k}$-point Hamiltonian at the next lithium concentration. As in the
$\mathbf{k}$-path transfer study, these calculations use four-qubit $e_g$ and
five-qubit $t_{2g}$ Hamiltonians, so the source and target calculations have
the same qubit-register size while differing in lithium composition.

Inter-composition transfer was compared using 100 sampled $\mathbf{k}$ points
for the four primary variational workflows.

\begin{figure}
\centering
\includegraphics[width=\linewidth]
{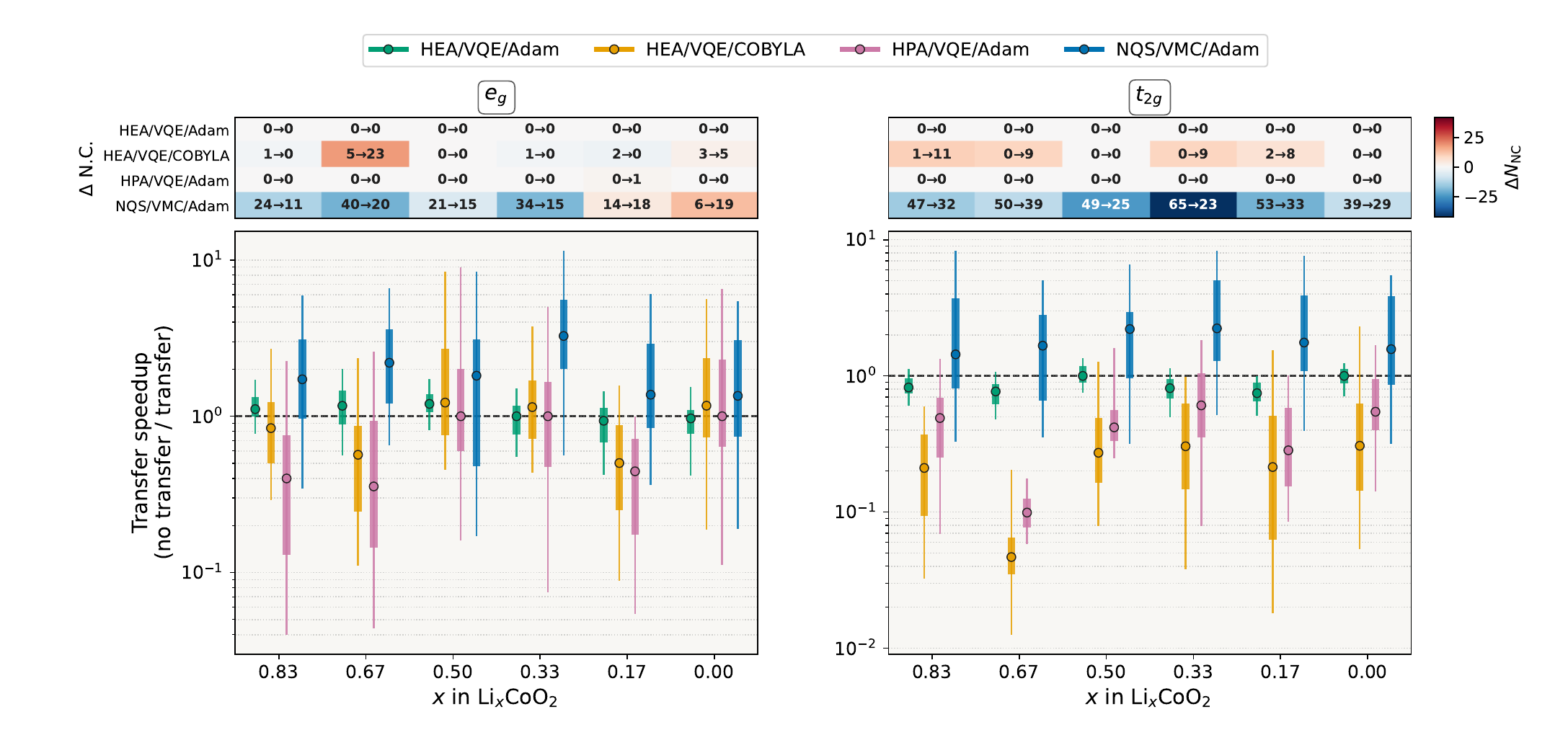}
\caption{\textbf{Effect of inter-composition parameter transfer on optimization
performance.}
Results are shown for the $e_g$ and $t_{2g}$ manifolds at six target lithium
concentrations using 100 sampled $\mathbf{k}$ points for HEA/VQE/Adam,
HEA/VQE/COBYLA, HPA/VQE/Adam, and NQS/VMC/Adam. For each method, manifold,
and composition, the paired transfer speedup is calculated at matching
$\mathbf{k}$ points as
$N_{\mathrm{steps}}^{\mathrm{no-transfer}}/
N_{\mathrm{steps}}^{\mathrm{transfer}}$.
Only $\mathbf{k}$ points that converged under both initialization strategies
are included in the paired speedup distributions; non-converged calculations
are reported separately in the upper N.C. strips. Circles indicate the median
paired speedup, thick vertical bars span the interquartile range, and thin bars
span the 5th--95th percentiles. The speedup axis is logarithmic, and the
horizontal dashed line at unity denotes identical optimization cost. Values
above unity indicate fewer optimization steps with inter-composition transfer,
whereas values below unity indicate slower convergence with transfer. In the
upper N.C. strips, each cell reports the number of non-converged $\mathbf{k}$
points as ``no transfer $\rightarrow$ inter-composition transfer.'' Cell color
represents
$\Delta N_{\mathrm{NC}} =
N_{\mathrm{NC,transfer}}-N_{\mathrm{NC,no\mbox{-}transfer}}$:
blue indicates fewer non-converged $\mathbf{k}$ points with transfer, red
indicates more, and white indicates no change. Non-converged points are
excluded from the paired speedup statistics but retained in the N.C. counts.}
\label{fig:intercomposition-transfer}
\end{figure}

Figure~\ref{fig:intercomposition-transfer} shows that the response to
inter-composition initialization is strongly workflow dependent.
NQS/VMC/Adam exhibits the clearest benefit, particularly for the $t_{2g}$
manifold, where the median paired speedup remains above unity and the number
of non-converged calculations is reduced across all six target compositions.
For $e_g$, the median speedup also remains above unity, although the effect on
convergence success becomes less uniform toward lower lithium concentrations.

In contrast, the VQE-based workflows show weaker or less consistent benefits
from direct inter-composition initialization. HEA/VQE/Adam remains close to
unity over much of the composition range, whereas HEA/VQE/COBYLA and
HPA/VQE/Adam frequently exhibit paired speedups below unity, indicating that
direct transfer across compositions can increase the number of optimization
steps for these workflows. The non-convergence statistics provide
complementary information: HEA/VQE/Adam and HPA/VQE/Adam remain largely
converged under both initialization strategies, whereas HEA/VQE/COBYLA shows
composition-dependent changes in convergence success.

These results demonstrate that transfer across lithium compositions is
strongly dependent on the variational workflow. In particular,
inter-composition initialization provides a consistently useful transfer route
for NQS/VMC/Adam, whereas the VQE-based workflows generally show stronger
benefits from neighboring-$\mathbf{k}$ transfer within a fixed composition,
as discussed in the preceding section.

As an additional full-mesh check, we repeated the HPA/VQE/Adam
inter-composition calculation over the complete $10\times10\times10$ mesh.
The qualitative transfer behavior remained consistent with that observed for
the 100 sampled $\mathbf{k}$ points, as shown in the Supporting Information.
The resulting direct band gap of 2.285~eV also remained close to the exact
embedded-Hamiltonian value of 2.240~eV and within the range of reported values
discussed above.

The distinct behavior of NQS/VMC/Adam is also noteworthy. Unlike the
VQE-based workflows, which favored sequential transfer between neighboring
$\mathbf{k}$-point Hamiltonians, NQS/VMC/Adam benefited from both parallel
initialization along the $\mathbf{k}$ path and transfer across neighboring
compositions. This broader tolerance to the choice of source Hamiltonian
suggests that the NQS parameterization may encode reusable information in a
form that is less dependent on a specific reciprocal-space trajectory or
composition. The present results are not sufficient to establish such
representation-level robustness generally, but they motivate further study
of the transferability of neural quantum states across more diverse
Hamiltonian families.

Overall, these results show that variational information can remain
transferable even when the lithium composition changes, although the benefit
depends strongly on the underlying workflow. We next examine whether this
transferability can be extended further to Hamiltonians of different
dimensions using AARTI.
\subsection{Cross-Size Transfer with AARTI}

The transfer strategies discussed above operate within fixed embedded Hamiltonian dimensions. We therefore asked whether transfer could remain effective when the Hamiltonian itself changes size. To address this question, we employed AARTI, in which Hamiltonian-aligned Pauli generators obtained from previously solved systems are reused to initialize larger target Hamiltonians.

Transfer was examined along the sequence of increasing supercell sizes,
$$2\times1\times1 \rightarrow 2\times2\times1 \rightarrow
4\times2\times1 \rightarrow 4\times4\times1$$, corresponding to
24, 48, 96, and 192 atoms, respectively. The $\mathbf{k}=0$
Hamiltonians of fully lithiated LiCoO$_2$ ($x=1.00$) were considered for
both the $e_g$ and $t_{2g}$ manifolds. Unlike the fixed-size parameter-transfer calculations described above, AARTI transfers variational information between Hamiltonians represented using different numbers of qubits. For the sequence considered here, each doubling of the supercell increases the embedded-Hamiltonian dimension by a factor of two and therefore increases the required qubit register by one qubit. Accordingly, the \(e_g\) calculations use 4, 5, 6, and 7 qubits for the \(2\times1\times1\), \(2\times2\times1\), \(4\times2\times1\), and \(4\times4\times1\) systems, respectively, while the corresponding \(t_{2g}\) calculations use 5, 6, 7, and 8 qubits. Three ansatz construction strategies were compared: a target-only baseline constructed entirely from the target Hamiltonian, a generator-only transfer scheme in which compatible source generators were reused while their transferred parameters were reset, and a full-transfer protocol in which both the Hamiltonian-aligned generators and their optimized source parameters were inherited from the previously converged calculation.
\begin{figure}
    \centering
    \includegraphics[width=0.95\linewidth]{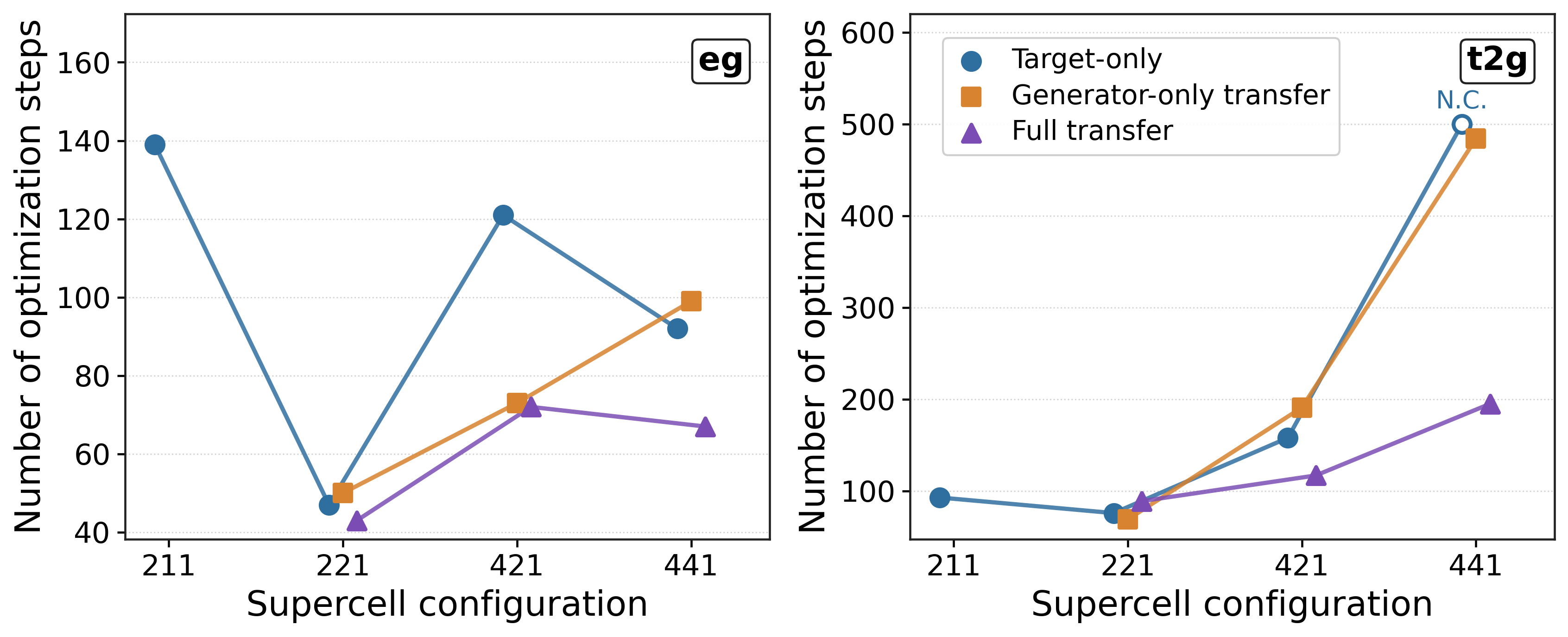}
\caption{\textbf{Cross-size transfer using AARTI.} The number of optimization steps required for convergence is shown along the supercell sequence $2\times1\times1 \rightarrow 2\times2\times1 \rightarrow 4\times2\times1 \rightarrow 4\times4\times1$ at $\mathbf{k}=0$ for the $e_g$ (left) and $t_{2g}$ (right) manifolds. Results are compared for target-only initialization, generator-only transfer, and full transfer. Generator-only transfer reuses compatible Hamiltonian-aligned Pauli generators while resetting their parameters, whereas full transfer reuses both the generator structure and optimized source parameters. The target-only calculation for the largest $t_{2g}$ Hamiltonian did not converge within 500 Adam steps, while full transfer converged successfully.}    \label{fig:aarti}
\end{figure}
Figure~\ref{fig:aarti} summarizes the resulting convergence behavior. For both orbital manifolds, the full-transfer protocol consistently reduced the number of optimization steps required for convergence relative to the target-only construction. Generator-only transfer also improved convergence, indicating that the inherited operator support and ordering themselves encode useful information about the target optimization landscape. However, the largest benefits were obtained when the optimized source parameters were transferred together with the generator structure.

For the $e_g$ manifold, the full-transfer calculations converged in 43, 72, and 67 optimization steps for the 221, 421, and 441 targets, respectively. The corresponding $t_{2g}$ calculations converged in 89, 117, and 195 steps. Notably, the target-only calculation for the largest $t_{2g}$ 441 Hamiltonian failed to satisfy the convergence criterion within the allotted 500 optimization steps, whereas the transferred ansatz converged successfully. This observation shows that Hamiltonian-aligned transfer can reduce optimization effort and can enable convergence within a fixed optimization budget for a larger target problem that did not converge under target-only initialization.

Taken together, these results demonstrate that transfer can extend beyond
neighboring $\mathbf{k}$ points and chemically related compositions to bridge
changing Hamiltonian dimensions. The ability to reuse both operator-level information and optimized parameters across scales provides a pathway toward applying variational quantum workflows to progressively larger condensed matter problems.

\section{Conclusion}

In this work, we developed a quantum-computing workflow for studying
defect-containing condensed-matter systems and applied it to
Li$_x$CoO$_2$ across multiple delithiation levels. By combining
first-principles downfolding, Wannierization, embedded tight-binding
Hamiltonians, variational quantum algorithms, and neural quantum states, we
constructed families of related Hamiltonians spanning reciprocal space,
composition, and increasing embedded system size. The resulting
electronic-structure calculations reproduce the non-rigid-band behavior
associated with lithium deintercalation.

Because these calculations require repeated optimization over many closely
related Hamiltonians, we systematically investigated whether previously
optimized variational information can be reused to accelerate subsequent
calculations. Parameter transfer reduced the optimization effort required to
satisfy a common convergence criterion while maintaining the common target
accuracy, with lower terminal errors observed in several cases. The effectiveness of transfer was
strongly workflow dependent. The VQE-based workflows generally benefited most
from serial transfer between neighboring $\mathbf{k}$ points, with
HPA/VQE/Adam exhibiting the largest convergence-step acceleration among the
methods considered. In contrast, NQS/VMC/Adam displayed a qualitatively
different transfer response, benefiting from parallel initialization along
the $\mathbf{k}$ path as well as transfer across neighboring lithium
compositions. This broader tolerance to the choice of source Hamiltonian
suggests that neural-state parameterizations may encode reusable information
differently from the circuit-based VQE ansatzes considered here.

The parameter-matched ansatz comparison further showed that the strong
transfer response of HPA/VQE/Adam cannot be explained solely by a smaller
number of variational parameters. The common transfer response observed for
Hamiltonian-aligned and low-Hamiltonian-weight Pauli ansatzes suggests that
shared multi-qubit Pauli-rotation parameterizations themselves may contribute
to parameter reusability, while the results are also consistent with an
additional contribution from Hamiltonian-informed generator selection. The persistence of the
inter-composition transfer trend for HPA/VQE/Adam when extending the
comparison from 100 selected $\mathbf{k}$ points to the complete
$10\times10\times10$ mesh further indicates that transfer behavior identified
from reduced benchmarks can remain representative at larger scale.

Finally, AARTI extends variational-information transfer beyond fixed
Hamiltonian dimensions by reusing compatible Hamiltonian-aligned operator
structure together with optimized source parameters. Across increasing
embedded system sizes, full transfer consistently reduced the optimization
effort relative to target-only initialization and enabled convergence for the
largest $t_{2g}$ case considered, for which the corresponding target-only
calculation did not satisfy the convergence criterion within the allotted
optimization budget.

Overall, these results show that transferability is not a universal property
of variational optimization, but depends on the representation, ansatz,
optimizer, and relationship between source and target Hamiltonians.
Identifying and exploiting these dependencies provides a practical route for
reducing repeated optimization cost in large Hamiltonian families and may
improve the scalability of quantum-computing workflows for realistic
materials simulations.
\section*{Acknowledgments}

The authors gratefully acknowledge Dr. Peter Mlkvik, Dr. Nicola Spaldin, and Dr. Marwa H. Farag
for their valuable guidance and insightful discussions during the course of
this research. 
The research was sponsored by the U.S. Department of Energy
(DOE) Office of Basic Energy Sciences, under grant number DESC0026309. This research used resources of the National Energy Research
Scientific Computing Center, a DOE Office of Science User Facility
supported by the Office of Science of the U.S. Department of Energy
under Contract No. DE-AC02-05CH11231 using NERSC award
NERSC DDR-ERCAP0036478.
M.S. would like to acknowledge the use of resources of the Oak Ridge Leadership Computing Facility at the Oak Ridge National Laboratory, which is supported by the Office of Science of the US Department of Energy under Contract No. DEAC05-00OR22725. This manuscript has in part been authored by UT-Battelle, LLC under Contract No. DE-AC05-00OR22725 with the U.S. Department of Energy.
The United States Government retains and the publisher, by accepting the article for publication, acknowledges that the US Government retains a nonexclusive, paid-up, irrevocable, worldwide license to publish or reproduce the published form of the manuscript, or allow others to do so, for US Government purposes. The Department of Energy will provide public access to these results of federally sponsored research in accordance with the DOE Public Access Plan (http://energy.gov/downloads/doe-publicaccess-plan).

\bibliographystyle{unsrt}
\bibliography{references}

\end{document}


\maketitle
\tableofcontents
\clearpage

\section{Reciprocal-Space Sampling}

Brillouin-zone integrations for the primary $2\times1\times1$ structures
employed a $4\times8\times2$ Monkhorst--Pack $\mathbf{k}$-point mesh.
Additional meshes were used for larger supercells and selected
electronic-structure calculations. Table~\ref{tab:kmesh_settings} summarizes
the reciprocal-space sampling used for the supercells considered in the
fixed-size and cross-size calculations.

\begin{table}[htbp]
\centering
\caption{Reciprocal-space sampling and relaxation settings used for each supercell.}
\label{tab:kmesh_settings}
\begin{tabular}{ccccc}
\toprule
Supercell & Number of atoms & Relaxation mesh & Electronic mesh & Purpose \\
\midrule
$2\times1\times1$ & 24  & $4\times8\times2$ & $4\times8\times2$ & Composition and dense-mesh studies \\
$2\times2\times1$ & 48  & $4\times4\times2$ & $4\times4\times2$ & Cross-size transfer \\
$4\times2\times1$ & 96  & $2\times4\times2$ & $2\times4\times2$ & Cross-size transfer \\
$4\times4\times1$ & 192 & $2\times2\times2$ & $2\times2\times2$ & Cross-size transfer \\
\bottomrule
\end{tabular}
\end{table}

\section{Wannier Downfolding and Embedded Hamiltonians}

Maximally localized Wannier functions were used to construct compact
Hamiltonian representations of the Co $3d$ manifold. The resulting Wannier
Hamiltonians were separated into the $t_{2g}$ and $e_g$ subspaces and
interpolated on the reciprocal-space grids used in the transfer studies.
The fixed-size $2\times1\times1$ embedded Hamiltonians have dimensions 16
and 32 for the $e_g$ and $t_{2g}$ manifolds, corresponding to four and five
qubits, respectively. Verified dimensions and Pauli-term counts available
for the cross-size calculations are summarized in
Table~\ref{tab:embedded_dimensions}. Values not directly verified from the
saved Hamiltonian datasets are intentionally not reported here.

\begin{table}[htbp]
\centering
\caption{Verified embedded-Hamiltonian dimensions and Pauli-term counts used in the fixed-size and cross-size calculations. A dash indicates that a Pauli-term count was not directly verified from the saved dataset used to prepare this Supporting Information.}
\label{tab:embedded_dimensions}
\begin{tabular}{ccccc}
\toprule
Supercell & Manifold & Matrix dimension & Number of qubits & Number of Pauli terms \\
\midrule
$2\times1\times1$ & $e_g$    & 16  & 4 & 136 \\
$2\times1\times1$ & $t_{2g}$ & 32  & 5 & 518 \\
$2\times2\times1$ & $e_g$    & 32  & 5 & 458 \\
$2\times2\times1$ & $t_{2g}$ & 64  & 6 & 1984 \\
$4\times2\times1$ & $e_g$    & 64  & 6 & 2072 \\
$4\times2\times1$ & $t_{2g}$ & 128 & 7 & 10677 \\
$4\times4\times1$ & $e_g$    & 128 & 7 & 8212 \\
$4\times4\times1$ & $t_{2g}$ & 256 & 8 & 32398 \\
\bottomrule
\end{tabular}
\end{table}

\section{Quantum Solver and Optimization Details}
\subsection{Optimization Statistics for k-Path Parameter Transfer}
\label{sec:si_kpath_statistics}

Table~\ref{tab:fig3_optimization_statistics} provides the numerical
statistics underlying the $\mathbf{k}$-path transfer comparison presented
in the main text. For each variational workflow, the $k_{000}$ Hamiltonian
was initialized independently and its convergence step is reported
separately. The distributions and non-convergence (N.C.) rates correspond
to the remaining 90 Hamiltonians, $k_{001}$--$k_{090}$, for which
no-transfer, serial-transfer, and parallel-transfer initialization were
compared. For the descriptive distributions in Table~\ref{tab:fig3_optimization_statistics}, non-converged calculations are excluded from the reported converged-sample mean, standard deviation, median, and quartiles, while their number and fraction are reported separately. The convergence-step speedups reported elsewhere use the convention defined in Sec.~\ref{sec:si_speedup}, in which non-converged calculations are counted at the corresponding maximum optimization-step budget.

\begin{table*}[htbp]
\centering
\caption{
Optimization-step statistics corresponding to the $\mathbf{k}$-path
transfer comparison in the main text. Statistics and non-convergence
(N.C.) rates correspond to $k_{001}$--$k_{090}$ ($N=90$).
The convergence step of the independently initialized $k_{000}$
calculation is reported separately and is not included in the
distributions.
}
\label{tab:fig3_optimization_statistics}
\scriptsize

\resizebox{\textwidth}{!}{%
\begin{tabular}{lllrrrrrr}
\toprule
Method &
Transfer &
Orbital &
Conv. &
N.C. &
Mean $\pm$ SD &
Median [Q1, Q3] &
N.C. (\%) &
$k_{000}$ step \\
\midrule

HEA/VQE/COBYLA & No transfer       & $e_g$    & 83/90 & 7  & 3235.3 $\pm$ 721.6  & 3199.0 [2752.5, 3639.0]   & 7.8  & 2341 \\
HEA/VQE/COBYLA & No transfer       & $t_{2g}$ & 58/90 & 32 & 14393.2 $\pm$ 3458.1 & 14720.5 [11646.8, 16846.8] & 35.6 & 8620 \\
HEA/VQE/COBYLA & Serial transfer   & $e_g$    & 85/90 & 5  & 668.5 $\pm$ 892.0   & 353.0 [249.0, 608.0]      & 5.6  & -- \\
HEA/VQE/COBYLA & Serial transfer   & $t_{2g}$ & 82/90 & 8  & 2895.7 $\pm$ 2937.5 & 2128.5 [866.5, 3223.8]    & 8.9  & -- \\
HEA/VQE/COBYLA & Parallel transfer & $e_g$    & 69/90 & 21 & 2751.9 $\pm$ 979.3  & 2466.0 [1981.0, 3427.0]   & 23.3 & -- \\
HEA/VQE/COBYLA & Parallel transfer & $t_{2g}$ & 81/90 & 9  & 11388.2 $\pm$ 6083.2 & 13071.0 [5410.0, 16549.0] & 10.0 & -- \\

\midrule

HEA/VQE/Adam & No transfer       & $e_g$    & 90/90 & 0 & 49.8 $\pm$ 7.4  & 49.0 [44.2, 53.0] & 0.0 & 49 \\
HEA/VQE/Adam & No transfer       & $t_{2g}$ & 90/90 & 0 & 64.8 $\pm$ 15.2 & 61.5 [53.2, 74.5] & 0.0 & 50 \\
HEA/VQE/Adam & Serial transfer   & $e_g$    & 90/90 & 0 & 15.9 $\pm$ 5.9  & 15.5 [11.0, 20.0] & 0.0 & -- \\
HEA/VQE/Adam & Serial transfer   & $t_{2g}$ & 90/90 & 0 & 29.7 $\pm$ 4.1  & 29.5 [26.0, 32.0] & 0.0 & -- \\
HEA/VQE/Adam & Parallel transfer & $e_g$    & 90/90 & 0 & 31.2 $\pm$ 9.5  & 28.5 [26.0, 31.8] & 0.0 & -- \\
HEA/VQE/Adam & Parallel transfer & $t_{2g}$ & 90/90 & 0 & 42.2 $\pm$ 13.3 & 36.0 [32.0, 47.0] & 0.0 & -- \\

\midrule

HPA/VQE/Adam & No transfer       & $e_g$    & 76/90 & 14 & 290.6 $\pm$ 88.2 & 257.0 [241.0, 357.2] & 15.6 & 242 \\
HPA/VQE/Adam & No transfer       & $t_{2g}$ & 90/90 & 0  & 248.0 $\pm$ 68.4 & 246.5 [182.2, 300.0] & 0.0  & 150 \\
HPA/VQE/Adam & Serial transfer   & $e_g$    & 90/90 & 0  & 11.2 $\pm$ 20.5 & 6.0 [4.0, 9.8] & 0.0 & -- \\
HPA/VQE/Adam & Serial transfer   & $t_{2g}$ & 90/90 & 0  & 12.4 $\pm$ 15.7 & 6.0 [6.0, 12.0] & 0.0 & -- \\
HPA/VQE/Adam & Parallel transfer & $e_g$    & 90/90 & 0  & 90.0 $\pm$ 81.3 & 61.0 [22.8, 148.5] & 0.0 & -- \\
HPA/VQE/Adam & Parallel transfer & $t_{2g}$ & 90/90 & 0  & 54.5 $\pm$ 33.0 & 51.0 [26.2, 82.8] & 0.0 & -- \\

\midrule

NQS/VMC/Adam & No transfer       & $e_g$    & 88/90 & 2  & 179.4 $\pm$ 71.6  & 154.5 [132.0, 211.2] & 2.2  & 135 \\
NQS/VMC/Adam & No transfer       & $t_{2g}$ & 45/90 & 45 & 504.9 $\pm$ 218.3 & 469.0 [360.0, 654.0] & 50.0 & 387 \\
NQS/VMC/Adam & Serial transfer   & $e_g$    & 75/90 & 15 & 46.1 $\pm$ 75.7   & 24.0 [15.0, 38.5] & 16.7 & -- \\
NQS/VMC/Adam & Serial transfer   & $t_{2g}$ & 81/90 & 9  & 36.7 $\pm$ 32.8   & 26.0 [19.0, 39.0] & 10.0 & -- \\
NQS/VMC/Adam & Parallel transfer & $e_g$    & 90/90 & 0  & 43.9 $\pm$ 55.4   & 21.5 [15.0, 51.0] & 0.0  & -- \\
NQS/VMC/Adam & Parallel transfer & $t_{2g}$ & 84/90 & 6  & 93.0 $\pm$ 122.1  & 54.5 [29.0, 90.2] & 6.7  & -- \\

\bottomrule
\end{tabular}%
}
\end{table*}

The statistics further show that transfer behavior depends on the
variational workflow and orbital manifold. Serial transfer produces the
lowest mean optimization cost for each of the VQE workflows. For
NQS/VMC/Adam, parallel transfer performs favorably for the $e_g$ manifold,
where all 90 target Hamiltonians converge, whereas serial transfer gives
the lower mean optimization cost for the $t_{2g}$ manifold. These
transfer trends were used to select the solver-specific transfer and
rescue strategies employed for the complete band-path calculations
reported in the main text.

\subsection{Parameter-Matched HEA--LPA--HPA Control}
\label{sec:si_parameter_matched_control}

The primary $\mathbf{k}$-path comparison in the main text compares complete
variational workflows and therefore includes differences in ansatz construction,
optimizer settings, and variational parameter count. In particular, the HPA
configuration used in the primary comparison contains 32 and 128 trainable
parameters for the $e_g$ and $t_{2g}$ manifolds, respectively, whereas the
corresponding HEA/VQE/Adam calculations contain 80 and 150 parameters.
To determine whether the strong serial-transfer response observed for HPA could
be explained primarily by this difference in variational dimension, we performed
a separate parameter-matched control calculation.

This control compares three VQE/Adam ansatzes:
\begin{enumerate}
    \item a hardware-efficient ansatz (HEA);
    \begin{figure}
        \centering
        \includegraphics[width=0.75\linewidth]{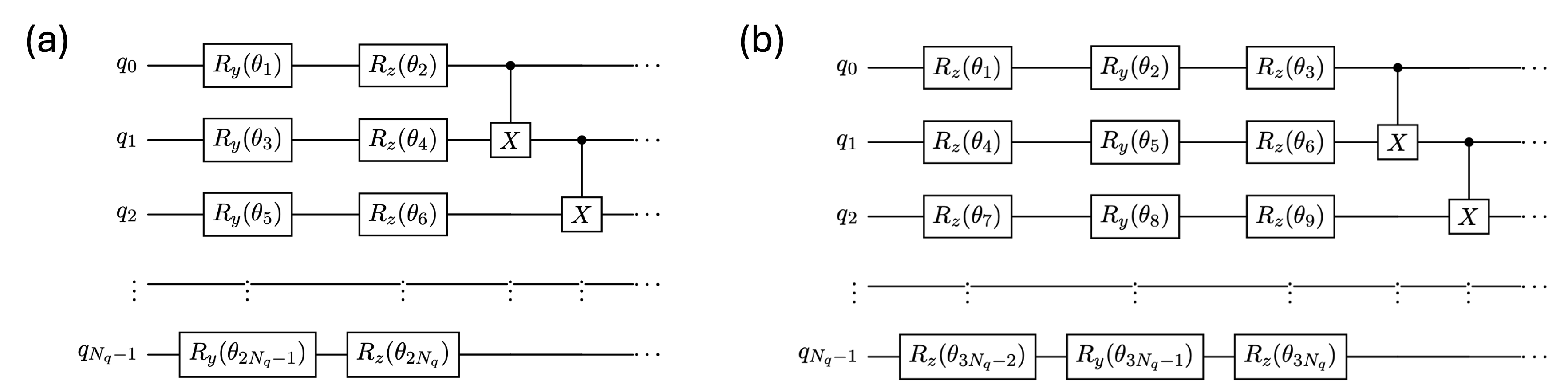}
        \label{fig:placeholder}
    \end{figure}
    \item the Hamiltonian-aligned Pauli ansatz (HPA); and
    \item a low-Hamiltonian-weight Pauli ansatz (LPA).
\end{enumerate}

The HPA and LPA control calculations are distinct from the lower-parameter HPA
configuration used in the primary four-workflow comparison. In the present
parameter-matched comparison, all three ansatzes contain 80 parameters for the
$e_g$ manifold and 150 parameters for the $t_{2g}$ manifold.

\begin{table}[htbp]
\centering
\caption{
Parameter counts and Pauli-generator support constraints used in the
parameter-matched HEA--LPA--HPA comparison. HPA and LPA use exactly the same
support distribution within each orbital manifold.
}
\label{tab:param_matched_ansatz}
\begin{tabular}{lccccc}
\toprule
Manifold &
Parameters &
1-qubit &
2-qubit &
3-qubit &
4-qubit \\
\midrule
$e_g$    & 80  & 5  & 19 & 36 & 20 \\
$t_{2g}$ & 150 & 10 & 50 & 90 & -- \\
\bottomrule
\end{tabular}
\end{table}

All calculations use the same case-10 Hamiltonians and the same 91-point
high-symmetry $\mathbf{k}$ path, $k_{000}$--$k_{090}$. The distributions shown
in the main-text parameter-matched comparison contain the 90 target Hamiltonians
$k_{001}$--$k_{090}$; the $k_{000}$ calculation is used for initialization and
is not included in the plotted distributions. All three approaches use VQE with
Adam and the same convergence criterion employed throughout the manuscript:
the variational energy must remain within 0.03~eV of the corresponding exact
eigenvalue for five consecutive optimization steps.

\subsubsection{HPA and LPA Generator Construction}

For the Pauli-based ansatzes, one fixed ordered generator list was constructed
from the $k_{000}$ Hamiltonian and retained across all 91 $\mathbf{k}$ points.
For a Hamiltonian written as
\begin{equation}
\hat{H}
=
\sum_j c_j \hat{P}_j,
\end{equation}
eligible non-identity Pauli generators were grouped according to their support,
defined as the number of qubits on which the Pauli string acts nontrivially.

HPA and LPA were constrained to contain exactly the same number of generators
within each support sector, as listed in
Table~\ref{tab:param_matched_ansatz}. The two ansatzes differ only in the
Hamiltonian-coefficient ranking used to choose generators within these
constraints. HPA preferentially selects eligible Pauli strings associated with
the largest values of $|c_j|$, whereas LPA reverses this ranking and selects
eligible strings associated with the smallest non-negligible values of $|c_j|$.
Identity terms and coefficients satisfying
\begin{equation}
|c_j| \leq 10^{-10}
\end{equation}
were excluded from both constructions.

The purpose of LPA is therefore not to define an independently optimized ansatz
architecture, but to provide a controlled Pauli-ansatz comparison in which the
number of parameters and the distribution of Pauli support are held fixed while
the Hamiltonian-weight criterion is changed.

\subsubsection{Initialization and Transfer Protocol}

HPA and LPA use the same random seed and the same dimension-matched initial
parameter vector at $k_{000}$. For no-transfer calculations, the same initial
parameter vector is reused independently for every target Hamiltonian,
\begin{equation}
\boldsymbol{\theta}^{(0)}_{k_n}
=
\boldsymbol{\theta}^{(0)}_{k_{000}},
\qquad
n=1,\ldots,90.
\end{equation}

For serial transfer, the optimized parameters obtained for the preceding
$\mathbf{k}$ point are used to initialize the next calculation,
\begin{equation}
\boldsymbol{\theta}^{(0)}_{k_n}
=
\boldsymbol{\theta}^{\ast}_{k_{n-1}},
\qquad
n=1,\ldots,90.
\end{equation}

Thus, the comparison directly probes how effectively a fixed variational
parameterization can reuse an optimized solution as the Hamiltonian changes
along the $\mathbf{k}$ path.

\subsubsection{Overlap Between HPA and LPA Generator Sets}

Matching the support distribution constrains the extent to which the HPA and
LPA generator sets can differ. For the $e_g$ manifold, HPA and LPA share
26 of the 80 selected Pauli generators, so that 54 generators differ between
the two ansatzes. For the $t_{2g}$ manifold, the two ansatzes share
100 of 150 generators.

The larger overlap for $t_{2g}$ follows directly from the imposed support
constraints. All 10 eligible one-qubit strings and all 50 eligible two-qubit
strings are required by the prescribed quotas. In addition, 90 three-qubit
generators are selected from a pool of 140 eligible strings, requiring an
overlap of at least 40 three-qubit terms between the high- and low-ranked
subsets. Consequently, 100 shared generators is the minimum possible overlap
under the present $t_{2g}$ construction.

This difference in overlap should be considered when comparing the separation
between HPA and LPA for the two orbital manifolds.

\subsubsection{Optimizer Settings and Interpretation}

The parameter-matched comparison controls the variational parameter count,
$\mathbf{k}$-point population, convergence criterion, and transfer definition.
For HPA and LPA, the Pauli-support distribution and initial parameter vector are
also controlled exactly.

The comparison between HEA and the Pauli ansatzes is not, however, a complete
architectural ablation. The existing HEA/VQE/Adam calculations use a learning
rate of 0.05 and a fixed-depth hardware-efficient circuit consisting of
10 layers, whereas the HPA and LPA calculations use a learning rate of 0.02
and Pauli-rotation circuits constructed from Hamiltonian-derived generators.
The HEA calculations use maximum iteration counts of 3000 for $e_g$ and
300 for $t_{2g}$, whereas the HPA and LPA control calculations use a maximum
of 500 optimization steps.

Accordingly, the HEA comparison is used primarily to test whether matching the
number of trainable parameters removes the strong HPA transfer response. It does
not by itself isolate every circuit-architectural contribution. The HPA--LPA
comparison provides the tighter control because the parameter count, Pauli
support distribution, initialization, optimizer settings, and transfer protocol
are matched, while the Hamiltonian-coefficient ranking used for generator
selection is varied.

The resulting behavior shows two trends. First, both HPA and LPA exhibit a
large reduction in optimization cost under serial transfer, indicating that the
transfer advantage is not determined solely by the selection of the
largest-magnitude Hamiltonian coefficients. Because the two ansatzes share the
same Pauli-support distribution, this common behavior is consistent with a
broader role for the Pauli-rotation parameterization and multi-qubit generator
structure in facilitating transfer between neighboring Hamiltonians. Second,
HPA retains the lowest serial-transfer optimization cost, suggesting that
Hamiltonian-informed generator selection provides an additional benefit beyond
this shared Pauli-ansatz effect.

The present control does not independently vary generator support, and therefore
does not by itself establish a causal dependence on Pauli weight or multi-qubit
support. Separating the effects of support, generator ordering, Hamiltonian
coefficient magnitude, and circuit structure provides a natural direction for
future work.

\subsection{Hamiltonian-Aligned Pauli VQE}
\label{sec:si_hpa}

The embedded Hamiltonian was decomposed as
\begin{equation}
\hat{H}
=
\sum_j c_j \hat{P}_j,
\end{equation}
where $\hat{P}_j$ denotes a Pauli string. A subset of non-identity Pauli
terms was selected to construct the variational state
\begin{equation}
|\psi(\boldsymbol{\theta})\rangle
=
\prod_{j=1}^{N_P}
\exp
\left(
-\frac{i}{2}\theta_j\hat{P}_j
\right)
|0\rangle^{\otimes N_q}.
\end{equation}

For fixed-size transfer calculations, the operator structure obtained from
the $\kpoint=000$ Hamiltonian of fully lithiated LiCoO$_2$ was retained,
while optimized parameters were transferred and reoptimized for related
Hamiltonians.

For the $e_g$ manifold, non-identity Pauli generators were ranked by
decreasing Hamiltonian coefficient magnitude, $|c_j|$. For the $t_{2g}$
manifold, generators were ranked using a low-support criterion: the pool was
restricted to strings with support at most three, diagonal $I/Z$ strings were
prioritized, and remaining strings were ordered by low support followed by
coefficient magnitude. No gradient estimate or commutation-based ranking was
used.

The maximum number of retained generators was 32, 64, 128, and 256 for the
$e_g$ 211, 221, 421, and 441 systems, respectively. For $t_{2g}$, the
corresponding limits were 128, 256, 384, and 512.

\subsection{NQS/VMC/Adam and qRBM-MCMC Sampling}
\label{sec:qrbm_mcmc}

As an alternative to the circuit-based VQE approaches, we employed a
neural-network quantum-state solver based on a quantum Restricted Boltzmann
Machine combined with variational Monte Carlo and Markov-chain Monte Carlo
sampling (qRBM-MCMC). The implementation follows the quantum-enabled
variational Monte Carlo framework introduced in Ref.~\cite{sajjan2026polynomially},
in which the variational wavefunction is represented by a Restricted
Boltzmann Machine (RBM), while a quantum-assisted proposal distribution is
used to improve exploration of the configurational space.

For a computational-basis configuration
$\mathbf{s}=(s_1,\ldots,s_n)$, with $s_i\in\{-1,+1\}$, the RBM
wavefunction is written as
\begin{equation}
\psi_{\boldsymbol{X}}(\mathbf{s})
=
\exp
\left(
\sum_i a_i s_i
\right)
\prod_j
\cosh
\left(
b_j+\sum_i W_{ij}s_i
\right),
\label{eq:qrbm_wavefunction}
\end{equation}
where
\begin{equation}
\boldsymbol{X}
=
\left(
\mathbf{a},
\mathbf{b},
\mathbf{W}
\right)
\end{equation}
collects the visible biases $a_i$, hidden biases $b_j$, and
visible--hidden coupling parameters $W_{ij}$. These parameters are varied
during optimization to minimize the expectation value of the embedded
Hamiltonian.

The probability distribution associated with the variational state is
\begin{equation}
p_{\boldsymbol{X}}(\mathbf{s})
=
\frac{
\left|
\psi_{\boldsymbol{X}}(\mathbf{s})
\right|^2
}{
\sum_{\mathbf{s}'}
\left|
\psi_{\boldsymbol{X}}(\mathbf{s}')
\right|^2
}.
\label{eq:qrbm_probability}
\end{equation}
Direct evaluation of the normalization in
Eq.~\eqref{eq:qrbm_probability} becomes impractical as the Hilbert-space
dimension increases. The qRBM-MCMC method therefore introduces a
tractable surrogate distribution that approximates the configurational
weight of the parent RBM and can be sampled through a Metropolis--Hastings
procedure.
For the second-order surrogate employed in the qRBM-MCMC construction,
the auxiliary distribution has the Ising-like form
\begin{equation}
\phi_{\boldsymbol{X}}(\mathbf{s})
\propto
\exp
\left[
\beta
\left(
\sum_i l_i(\boldsymbol{X})s_i
+
\sum_{i<j}
J_{ij}(\boldsymbol{X})s_i s_j
\right)
\right],
\label{eq:qrbm_surrogate}
\end{equation}
where $l_i(\boldsymbol{X})$ and $J_{ij}(\boldsymbol{X})$ are effective
one- and two-body parameters determined from the instantaneous RBM
parameters. The original configurational distribution can be represented
in terms of this surrogate through a configuration-dependent correction
factor,
\begin{equation}
p_{\boldsymbol{X}}(\mathbf{s})
\propto
\kappa(\mathbf{s};\boldsymbol{X})
\phi_{\boldsymbol{X}}(\mathbf{s}),
\label{eq:qrbm_factorization}
\end{equation}
with $\kappa(\mathbf{s};\boldsymbol{X})$ accounting for the difference
between the parent and surrogate distributions. Further details of the
surrogate construction and its polynomial approximation are given in
Ref.~\cite{sajjan2026polynomially}.

Samples from Eq.~\eqref{eq:qrbm_surrogate} are generated using a
Metropolis--Hastings Markov chain. Given a current configuration
$\mathbf{s}^{(r)}$ and a proposed configuration
$\mathbf{s}^{(r+1)}$, the acceptance probability is
\begin{equation}
A\left(
\mathbf{s}^{(r)}
\rightarrow
\mathbf{s}^{(r+1)}
\right)
=
\min
\left[
1,
\frac{
\phi_{\boldsymbol{X}}
\left(
\mathbf{s}^{(r+1)}
\right)
P_{\mathrm{prop}}
\left(
\mathbf{s}^{(r)}
\middle|
\mathbf{s}^{(r+1)}
\right)
}{
\phi_{\boldsymbol{X}}
\left(
\mathbf{s}^{(r)}
\right)
P_{\mathrm{prop}}
\left(
\mathbf{s}^{(r+1)}
\middle|
\mathbf{s}^{(r)}
\right)
}
\right].
\label{eq:qrbm_mh}
\end{equation}
Because only ratios of surrogate probabilities enter
Eq.~\eqref{eq:qrbm_mh}, the normalization constant of
$\phi_{\boldsymbol{X}}$ is not required explicitly.

In the quantum-enhanced implementation, candidate configurations are
generated using a parameterized quantum evolution tailored to the
surrogate Hamiltonian. Defining
\begin{equation}
\hat{h}_1
=
\sum_i
l_i(\boldsymbol{X})\hat{Z}_i
+
\sum_{i<j}
J_{ij}(\boldsymbol{X})
\hat{Z}_i\hat{Z}_j
\end{equation}
and the transverse mixing Hamiltonian
\begin{equation}
\hat{h}_2
=
\sum_i \hat{X}_i,
\end{equation}
the proposal unitary is constructed from
\begin{equation}
U(\tau,\gamma)
=
\exp
\left[
i\tau
\left(
\gamma\hat{h}_1
+
(1-\gamma)\hat{h}_2
\right)
\right].
\label{eq:qrbm_unitary}
\end{equation}
For an incumbent configuration $\mathbf{s}^{(r)}$, the corresponding
proposal probability is
\begin{equation}
P_{\mathrm{prop}}
\left(
\mathbf{s}^{(r+1)}
\middle|
\mathbf{s}^{(r)}
\right)
=
\left|
\left\langle
\mathbf{s}^{(r+1)}
\right|
U(\tau,\gamma)
\left|
\mathbf{s}^{(r)}
\right\rangle
\right|^2.
\label{eq:qrbm_proposal}
\end{equation}
In practice, the evolution operator is implemented using a product-formula
(Trotterized) circuit, so that the quantum circuit acts only as a sampler
for candidate bit strings; the RBM itself remains a classically stored and
optimized neural-network quantum state. The circuit is measured in the
computational basis, and the resulting bit string is passed to the
Metropolis--Hastings acceptance step.

After an initial burn-in period, the retained configurations form a sample
set
\begin{equation}
\mathcal{S}
=
\left\{
\mathbf{s}^{(r)}
\right\}_{r=1}^{N_s},
\end{equation}
distributed according to the surrogate model. For an embedded Hamiltonian
$\hat{H}$, the local energy associated with a configuration is
\begin{equation}
E_{\mathrm{loc}}
\left(
\mathbf{s};
\boldsymbol{X}
\right)
=
\frac{
\displaystyle
\sum_{\mathbf{s}'}
\langle\mathbf{s}|\hat{H}|\mathbf{s}'\rangle
\psi_{\boldsymbol{X}}(\mathbf{s}')
}{
\psi_{\boldsymbol{X}}(\mathbf{s})
}.
\label{eq:qrbm_local_energy}
\end{equation}
The variational energy is then estimated from the sampled configurations
using the corresponding correction factors,
\begin{equation}
\widehat{E}(\boldsymbol{X})
=
\frac{
\displaystyle
\sum_{\mathbf{s}\in\mathcal{S}}
\kappa(\mathbf{s};\boldsymbol{X})
E_{\mathrm{loc}}
\left(
\mathbf{s};\boldsymbol{X}
\right)
}{
\displaystyle
\sum_{\mathbf{s}\in\mathcal{S}}
\kappa(\mathbf{s};\boldsymbol{X})
}.
\label{eq:qrbm_energy_estimator}
\end{equation}
The same sampled configurations are used to construct stochastic
estimators of the derivatives of the variational energy with respect to
the RBM parameters. These gradients are subsequently supplied to the
classical optimizer to update
$\mathbf{a}$, $\mathbf{b}$, and $\mathbf{W}$, and the sampling and
optimization procedure is repeated until the convergence criterion
described below is satisfied.

For the fixed-size Hamiltonians considered in this work, the $e_g$
qRBM-MCMC calculations used four qubits and quantum sampling circuits
containing 28 parameterized gates, whereas the $t_{2g}$ calculations used
five qubits and 34 parameterized gates. Parameter transfer between related
Hamiltonians was performed by reusing the fully optimized RBM parameter
vector,
\begin{equation}
\boldsymbol{X}^{(0)}_{\mathrm{target}}
=
\boldsymbol{X}^{\ast}_{\mathrm{source}},
\label{eq:qrbm_transfer}
\end{equation}
so that all visible biases, hidden biases, and coupling parameters were
transferred together. Subsequent optimization was then carried out for the
target Hamiltonian using the same qRBM-MCMC sampling and stochastic
optimization procedure.

The detailed derivation of the surrogate model, construction of the
quantum-enhanced proposal distribution, convergence properties of the
Markov chain, and resource scaling of the qRBM-MCMC algorithm are given
in Ref.~\cite{sajjan2026polynomially}. The present work uses this method as a
variational eigensolver and focuses on the extent to which previously
optimized qRBM parameters can be transferred between closely related
embedded Hamiltonians.

\subsection{Solver Hyperparameters}

For the Hamiltonian-aligned Pauli calculations, variational parameters were
optimized using Adam with parameter-shift gradients. All AARTI and cross-size
Pauli-transfer calculations used a maximum of 500 optimization steps, a
convergence threshold of 0.03~eV, and a patience window of five consecutive
iterations. The $e_g$ calculations used a learning rate of 0.02, whereas the
$t_{2g}$ calculations used a learning rate of 0.01.

New Pauli-rotation parameters were initialized from the corresponding
Hamiltonian coefficient according to
\begin{equation}
\theta_m^{(0)} = 0.25\,
\frac{\mathrm{Re}(c_m)}{\max_j |c_j|}.
\end{equation}
For the $t_{2g}$ $2\times1\times1$ source calculation, a small warm-start
jitter of 0.05 was used; the remaining redo calculations used deterministic
coefficient-based initialization. All reported AARTI redo calculations used
seed 1.

\begin{table}[htbp]
\centering
\caption{Optimizer settings used for HPA/VQE/Adam and AARTI cross-size calculations.}
\label{tab:solver_settings}
\resizebox{\linewidth}{!}{%
\begin{tabular}{lcccccc}
\toprule
Method & Manifold & Optimizer & Learning rate & Maximum steps & Initial parameter rule & Number of seeds \\
\midrule
HPA/VQE/Adam & $e_g$ & Adam & 0.02 & 500 & $0.25\,\mathrm{Re}(c_m)/\max_j|c_j|$ & 1 \\
HPA/VQE/Adam & $t_{2g}$ & Adam & 0.01 & 500 & coefficient based; 0.05 jitter for source & 1 \\
\bottomrule
\end{tabular}}
\end{table}

\subsection{Convergence Criterion and Speedup Definition}
\label{sec:si_speedup}

Optimization was considered converged when the variational energy was within
0.03~eV of the corresponding exact eigenvalue for five consecutive
optimization steps. The convergence-step speedup was calculated as
\begin{equation}
S_{\mathrm{iter}}=
\frac{\left\langle N_{\mathrm{steps}}^{\mathrm{no-transfer}}\right\rangle}
{\left\langle N_{\mathrm{steps}}^{\mathrm{transfer}}\right\rangle},
\end{equation}
where averages were evaluated over the same set of target Hamiltonians. For
speedup calculations, a target that did not satisfy the convergence criterion
within the allotted budget was counted at the maximum number of optimization
steps for that run. This convention is distinct from the descriptive
converged-sample statistics in Table~\ref{tab:fig3_optimization_statistics},
where non-converged calculations are reported separately rather than included
in the mean, standard deviation, median, or quartiles.

\section{Transfer Protocols}

\subsection{Serial and Parallel Transfer Across \texorpdfstring{$\mathbf{k}$}{k}-Points}

For serial transfer, the optimized parameter vector from one $\mathbf{k}$
point was used to initialize the next point along the chosen traversal,
\begin{equation}
\boldsymbol{\theta}^{(0)}_{\mathbf{k}_{i+1}}
=\boldsymbol{\theta}^{\ast}_{\mathbf{k}_i}.
\end{equation}
Along the high-symmetry $\Gamma$--$M$--$K$--$\Gamma$ path, $k_{000}$ was
initialized independently and the optimized parameters were propagated
sequentially through $k_{001}$--$k_{090}$. In parallel transfer, parameters
from a common converged reference calculation were distributed to multiple
target Hamiltonians without sequential propagation.

\subsection{Selected k-Path Workflows and Rescue-Aware Step Accounting}
\label{sec:si_selected_workflow_accounting}

The transfer preferences observed across the 90 target Hamiltonians were used
to select one workflow-specific transfer strategy for completing the
$\mathbf{k}$ path. HEA/VQE/Adam and HPA/VQE/Adam used serial neighboring-
$\mathbf{k}$ transfer without additional rescue. HEA/VQE/COBYLA used serial
transfer as the primary strategy, with unconverged targets subsequently
reinitialized from a previously converged neighboring $\mathbf{k}$ point.
NQS/VMC/Adam used parallel transfer as the common primary strategy across the
two manifolds, with unconverged $t_{2g}$ targets subsequently reinitialized
from successfully converged neighboring $\mathbf{k}$ points.

For the workflow-level speedups reported in the main text, the cold-start cost
for each target was defined as
\begin{equation}
C_k^{\mathrm{cold}}=
\begin{cases}
s_k^{\mathrm{cold}}, & \text{if the cold-start calculation converged},\\
L_k^{\mathrm{cold}}, & \text{otherwise},
\end{cases}
\end{equation}
where $s_k$ is the recorded convergence step and $L_k$ is the configured
maximum optimization-step limit. The selected-workflow cost was defined as
\begin{equation}
C_k^{\mathrm{selected}}=
\begin{cases}
s_k^{\mathrm{primary}}, & \text{if the primary transfer converged},\\
L_k^{\mathrm{primary}}+s_k^{\mathrm{rescue}},
& \text{if a rescue calculation was required}.
\end{cases}
\end{equation}
The workflow-level convergence-step speedup was then evaluated over the same
90 transferred targets as
\begin{equation}
S_{\mathrm{workflow}}
=
\frac{\frac{1}{90}\sum_{k=1}^{90}C_k^{\mathrm{cold}}}
{\frac{1}{90}\sum_{k=1}^{90}C_k^{\mathrm{selected}}}.
\end{equation}
Thus, the reported quantity is a ratio of mean workflow costs rather than a
mean of pointwise speedup ratios.

For NQS/VMC/Adam, parallel transfer was used as the primary strategy; no
$e_g$ targets and six $t_{2g}$ targets required neighboring-$\mathbf{k}$
rescue. The corresponding maximum step limits were 500 and 1000 steps for
$e_g$ and $t_{2g}$, respectively. For HEA/VQE/COBYLA, serial transfer was
used as the primary strategy; five $e_g$ and eight $t_{2g}$ targets required
backward-neighbor rescue, with maximum step limits of 5000 and 20000 steps,
respectively. HEA/VQE/Adam used serial transfer and required no rescue; no
cold-start or transfer failures occurred. HPA/VQE/Adam also used serial
transfer without rescue; fourteen cold-start $e_g$ targets failed and were
therefore charged the 500-step limit in the cold-start accounting. Every
selected workflow ultimately converged at all 90 transferred targets.

\begin{table}[htbp]
\centering
\caption{Workflow-level convergence-step speedups for the selected
solver-specific $\mathbf{k}$-path transfer workflows. All values are computed
over $k_{001}$--$k_{090}$ using the rescue-aware accounting defined in
Sec.~\ref{sec:si_selected_workflow_accounting}.}
\label{tab:si_selected_workflow_speedups}
\begin{tabular}{lccc}
\toprule
Method & Selected transfer workflow & $e_g$ speedup & $t_{2g}$ speedup \\
\midrule
NQS/VMC/Adam
& Parallel + neighboring-$\mathbf{k}$ rescue
& $4.25\times$ & $4.85\times$ \\
HEA/VQE/COBYLA
& Serial + backward-neighbor rescue
& $3.65\times$ & $3.67\times$ \\
HEA/VQE/Adam
& Serial neighboring-$\mathbf{k}$ transfer
& $3.13\times$ & $2.18\times$ \\
HPA/VQE/Adam
& Serial neighboring-$\mathbf{k}$ transfer
& $28.74\times$ & $19.90\times$ \\
\bottomrule
\end{tabular}
\end{table}

\subsection{Inter-Composition Transfer Across Lithium Concentrations}
\label{sec:si_intercomposition}

Inter-composition transfer maps optimized parameters from a Hamiltonian at one
lithium concentration to the corresponding $\mathbf{k}$-point Hamiltonian at
the next composition,
\begin{equation}
\boldsymbol{\theta}^{(0)}_{x_{j+1},\mathbf{k}_i}
=
\boldsymbol{\theta}^{\ast}_{x_j,\mathbf{k}_i}.
\end{equation}
Transfer was performed sequentially along the delithiation direction,
\begin{equation}
1.00 \rightarrow 0.83 \rightarrow 0.67 \rightarrow 0.50
\rightarrow 0.33 \rightarrow 0.17 \rightarrow 0.00.
\end{equation}
The comparison reported in the main text used 100 sampled $\mathbf{k}$ points
from the complete $10\times10\times10$ mesh for each of the four primary
variational workflows.

For the paired distributions shown in the main-text inter-composition figure,
the transfer speedup at a matching $\mathbf{k}$ point is
\begin{equation}
S_{\mathrm{paired}}(\mathbf{k})
=
\frac{N_{\mathrm{steps}}^{\mathrm{no-transfer}}(\mathbf{k})}
{N_{\mathrm{steps}}^{\mathrm{transfer}}(\mathbf{k})}.
\end{equation}
Only $\mathbf{k}$ points that converged under both initialization strategies
are included in these paired speedup distributions. Non-converged
calculations are instead reported separately through the N.C. counts. This
paired statistic is therefore distinct from the workflow-level ratio of mean
costs used for the selected $\mathbf{k}$-path workflows.

\begin{figure}[htbp]
    \centering
    \includegraphics[width=\linewidth]
    {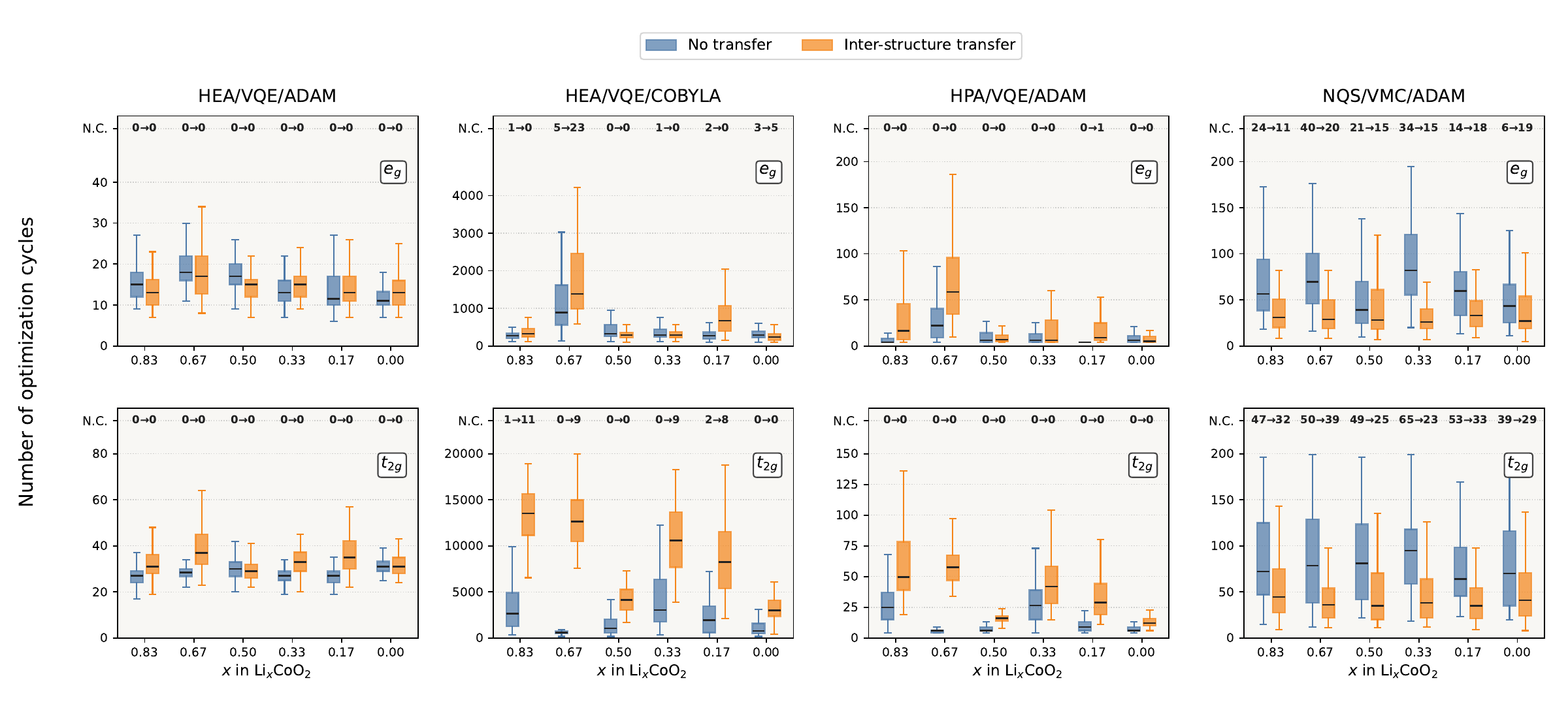}
    \caption{
    Inter-composition optimization-step distributions for the 100 sampled
    $\mathbf{k}$ points. Results are shown for HEA/VQE/Adam,
    HEA/VQE/COBYLA, HPA/VQE/Adam, and NQS/VMC/Adam for the $e_g$ and
    $t_{2g}$ manifolds across the six target lithium concentrations.
    No-transfer and inter-composition-transfer calculations are shown
    separately, providing the absolute optimization-step distributions
    underlying the paired-speedup comparison in the main text.
    }
    \label{fig:si_intercomposition_100k_boxplots}
\end{figure}

Figure~\ref{fig:si_intercomposition_100k_boxplots} shows the underlying
optimization-step distributions with and without inter-composition transfer
for the 100 sampled $\mathbf{k}$ points. These distributions complement the
paired-speedup representation in the main text by showing the absolute
optimization-step behavior for each workflow and orbital manifold.

\subsection{Full-Mesh HPA/VQE/Adam Inter-Composition Check}
\label{sec:si_fullmesh_intercomposition}

As an additional check of whether the behavior observed from the 100 sampled
$\mathbf{k}$ points remains representative of the complete reciprocal-space
mesh, the HPA/VQE/Adam inter-composition calculation was repeated for all
1000 $\mathbf{k}$ points of the $10\times10\times10$ mesh. This calculation
is used as a full-mesh consistency check rather than as a computational
scaling analysis.

\begin{figure}[htbp]
    \centering
    \includegraphics[width=\linewidth]
    {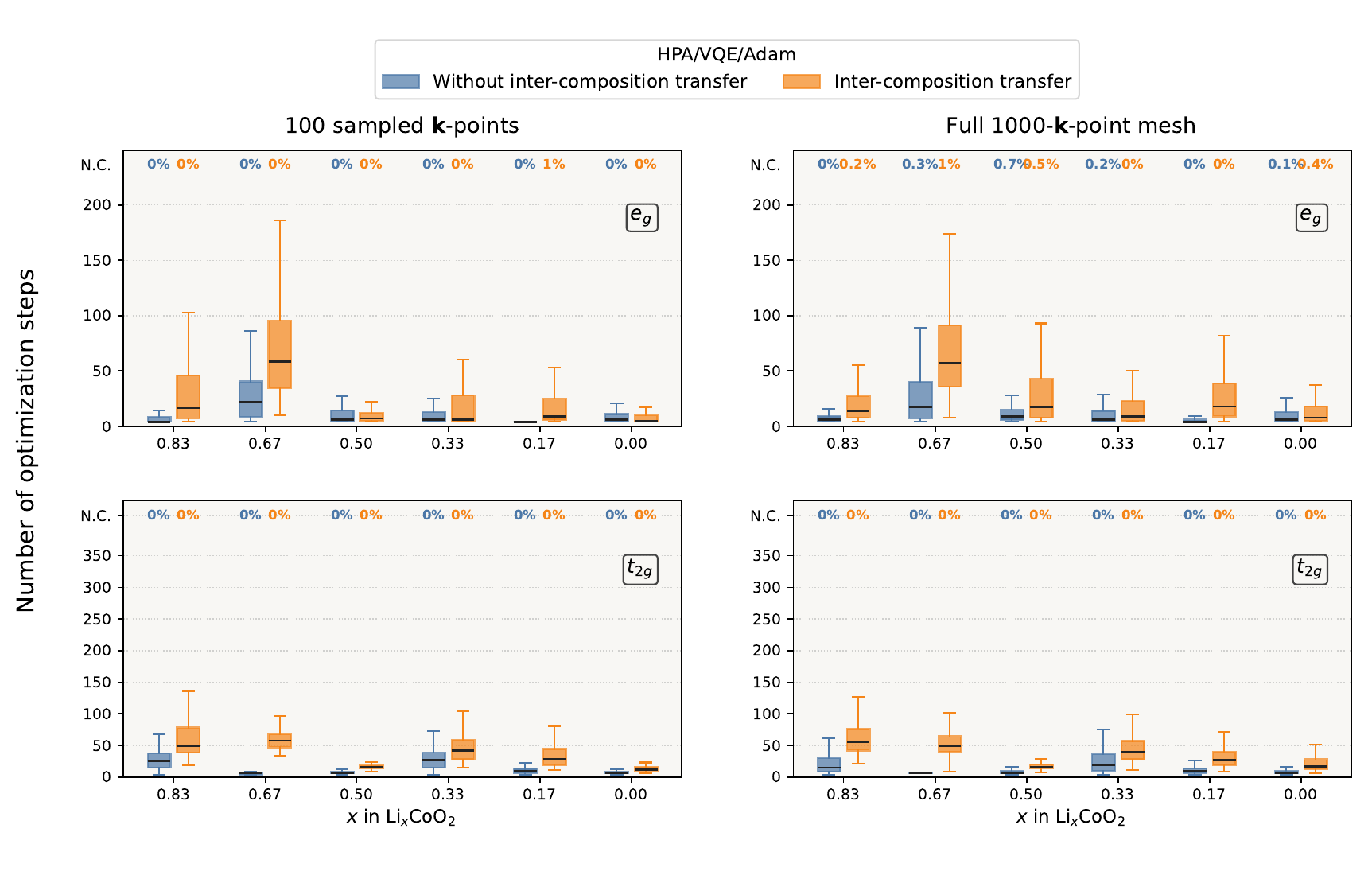}
    \caption{
    Comparison of HPA/VQE/Adam inter-composition optimization behavior for
    100 sampled $\mathbf{k}$ points and the complete
    $10\times10\times10$ mesh containing 1000 $\mathbf{k}$ points.
    The upper and lower rows correspond to the $e_g$ and $t_{2g}$ manifolds,
    respectively. Distributions are shown with and without inter-composition
    transfer at each target lithium concentration, with non-convergence
    fractions reported above the distributions. The similar qualitative
    behavior in the sampled and full-mesh calculations supports the
    representativeness of the 100-point comparison for this system.
    }
    \label{fig:si_hpa_100_vs_1000}
\end{figure}

As shown in Fig.~\ref{fig:si_hpa_100_vs_1000}, the composition-dependent
optimization-step trends obtained from the sampled and full-mesh calculations
remain qualitatively similar for both orbital manifolds. The full-mesh
calculation also retained the expected electronic-structure behavior. The
exact embedded-Hamiltonian direct and indirect gaps were 2.2401 and
2.1945~eV, respectively, while the corresponding HPA/VQE/Adam values were
2.2847 and 2.2352~eV. The direct-gap value therefore remains close to the
exact embedded-Hamiltonian result and within the range of reported values
discussed in the main text.

\begin{table}[htbp]
\centering
\caption{Electronic-structure quantities from the complete
$10\times10\times10$ mesh for the exact embedded Hamiltonian and the final
HPA/VQE/Adam calculation.}
\label{tab:si_fullmesh_gaps}
\begin{tabular}{lcc}
\toprule
Quantity & Exact & HPA/VQE/Adam \\
\midrule
Direct gap (eV) & 2.2401 & 2.2847 \\
Indirect gap (eV) & 2.1945 & 2.2352 \\
CBM $\mathbf{k}$ point & $k_{051}$ & $k_{051}$ \\
VBM $\mathbf{k}$ point & $k_{294}$ & $k_{294}$ \\
\bottomrule
\end{tabular}
\end{table}

\subsection{Transferred Variational Information}

For the fixed-size $\mathbf{k}$-path calculations, the complete available
variational parameter object was transferred rather than a selected subset.
For HEA/VQE/Adam and HEA/VQE/COBYLA this corresponds to all 80 parameters for
$e_g$ and all 150 parameters for $t_{2g}$. For HPA/VQE/Adam, all 32 and 128
Pauli-rotation parameters were transferred for $e_g$ and $t_{2g}$,
respectively. For NQS/VMC/Adam, the transferred object was the fully optimized
RBM parameter vector
$\boldsymbol{X}=(\mathbf{a},\mathbf{b},\mathbf{W})$. The numbers 28 and 34
reported for the NQS sampling circuits refer to parameterized quantum gates in
the four- and five-qubit sampling circuits, respectively, and are therefore
not used here as counts of transferred RBM variational parameters.

\begin{table}[htbp]
\centering
\caption{Variational information transferred in the fixed-size
$\mathbf{k}$-path calculations.}
\label{tab:transfer_parameter_counts}
\begin{tabular}{lccc}
\toprule
Method & $e_g$ & $t_{2g}$ & Transferred object \\
\midrule
HEA/VQE/Adam & 80 parameters & 150 parameters & Full HEA parameter vector \\
HEA/VQE/COBYLA & 80 parameters & 150 parameters & Full HEA parameter vector \\
HPA/VQE/Adam & 32 parameters & 128 parameters & Full Pauli-rotation parameter vector \\
NQS/VMC/Adam & -- & -- & Full RBM vector $\boldsymbol{X}$ \\
\bottomrule
\end{tabular}
\end{table}

\section{AARTI Cross-Size Transfer}

\subsection{Augmented Ansatz Reuse for Target Initialization}

Augmented Ansatz Reuse for Target Initialization (AARTI) transfers
Hamiltonian-aligned Pauli ansatz information between embedded Hamiltonians of
different sizes. Compatible Pauli-rotation generators from a previously solved
source Hamiltonian are embedded into the larger target register, retained when
they are present in the target Hamiltonian Pauli pool, and augmented with
target-specific generators. In full transfer, optimized parameters associated
with the retained source generators are also inherited; in generator-only
transfer, the compatible generator structure is retained while the transferred
parameters are reset. The detailed matching and augmentation rules are given
below.

\subsection{Cross-Size Transfer Sequence}

Cross-size transfer was investigated along the sequence
\begin{equation}
2\times1\times1
\rightarrow
2\times2\times1
\rightarrow
4\times2\times1
\rightarrow
4\times4\times1
\end{equation}
at $\kpoint=\mathbf{0}$ for the $e_g$ and $t_{2g}$ manifolds.

Three initialization strategies were compared:

\begin{enumerate}[label=(\roman*)]
    \item \textbf{Target-only}: the ansatz was constructed only from the
    target Hamiltonian.

    \item \textbf{Generator-only transfer}: compatible source generators
    were retained, but their parameters were reset.

    \item \textbf{Full transfer}: compatible source generators and their
    optimized source parameters were both retained.
\end{enumerate}

\subsection{Generator Matching and Augmentation}

Generator matching in AARTI was performed using exact Pauli-string
compatibility with the target Hamiltonian. For each source-target pair, the
optimized source ansatz supplied a list of source Pauli generators
$\{P_i^{(s)}\}$. Each source generator was embedded into the larger target
qubit register by identity padding,
\begin{equation}
P_i^{(s)}
\rightarrow
\tilde{P}_i
=
I^{\otimes \Delta N_q}\otimes P_i^{(s)}
\quad \mathrm{or} \quad
\tilde{P}_i
=
P_i^{(s)}\otimes I^{\otimes \Delta N_q},
\end{equation}
where $\Delta N_q=N_q^{(t)}-N_q^{(s)}$.

When \texttt{pad\_side=auto}, both padding directions were tested. The
direction was chosen by counting exact matches between the padded source
strings and the target Hamiltonian Pauli support,
\begin{equation}
N_{\mathrm{left}}
=
\left|
\left\{
I^{\otimes \Delta N_q}P_i^{(s)}
\in \mathcal{P}(H_t)
\right\}
\right|,
\qquad
N_{\mathrm{right}}
=
\left|
\left\{
P_i^{(s)}I^{\otimes \Delta N_q}
\in \mathcal{P}(H_t)
\right\}
\right|.
\end{equation}
Left padding was used when
$N_{\mathrm{left}}\geq N_{\mathrm{right}}$; otherwise right padding was used.

After the padding direction was selected, a padded source generator was
retained only if the exact padded Pauli string appeared in the target
Hamiltonian Pauli pool,
\begin{equation}
\tilde{P}_i\in\mathcal{P}(H_t).
\end{equation}
Source generators that failed this exact-match criterion were discarded. The
retained generators were inserted into the target ansatz in their original
source order.

The remaining ansatz slots were filled using target-specific Pauli strings not
already inherited from the source,
\begin{equation}
Q_j \in
\mathcal{P}(H_t)\setminus\mathcal{G}_{\mathrm{kept}}.
\end{equation}
For the $e_g$ manifold, candidate target generators were ranked by decreasing
coefficient magnitude. For the $t_{2g}$ manifold, the candidate pool was
restricted to Pauli strings with support at most three; diagonal $I/Z$ strings
were prioritized, followed by lower-support strings and then coefficient
magnitude.

The final target ansatz was therefore
\begin{equation}
\mathcal{G}^{(t)}
=
\mathcal{G}_{\mathrm{kept}}
\cup
\mathcal{G}_{\mathrm{new}},
\end{equation}
where $\mathcal{G}_{\mathrm{kept}}$ contains compatible padded source
generators and $\mathcal{G}_{\mathrm{new}}$ contains newly added target
Hamiltonian generators. In generator-only transfer, the retained source
generators were reused with their angles reset to zero. In full transfer, the
retained generators were initialized with the optimized source angles.

In the cross-size AARTI calculations reported here, \texttt{pad\_side=auto}
was used for all source-data transfer runs. For generator-only transfer,
\texttt{source\_angle\_scale=0.0}; for full transfer,
\texttt{source\_angle\_scale=1.0}. Newly added target generators were not
frozen during optimization, corresponding to
\texttt{freeze\_new\_term\_steps=0}. For the $e_g$ manifold, Pauli terms were
ranked using \texttt{pool\_order=magnitude}. For the $t_{2g}$ manifold, terms
were ranked using \texttt{pool\_order=low\_support\_then\_magnitude} with
\texttt{max\_support=3} and \texttt{diagonal\_first=true}.

\subsection{Generator Budgets in the AARTI Calculations}

The total Pauli-generator budgets were 32, 64, 128, and 256 for the $e_g$
$2\times1\times1$, $2\times2\times1$, $4\times2\times1$, and
$4\times4\times1$ systems, respectively, and 128, 256, 384, and 512 for the
corresponding $t_{2g}$ systems. In the full-transfer AARTI runs, the numbers of
retained source generators and newly added target generators were 32+32,
63+65, and 126+130 for the $e_g$ $2\times2\times1$, $4\times2\times1$, and
$4\times4\times1$ targets, respectively. For $t_{2g}$, the corresponding
retained+new counts were 128+128, 254+130, and 360+152.

\section{Electronic-Structure Evolution Across Lithium Concentration}
\label{sec:si_electronic_structure}

\begin{figure}[htbp]
    \centering
    \begin{subfigure}[t]{0.49\linewidth}
        \centering
        \includegraphics[width=\linewidth]
        {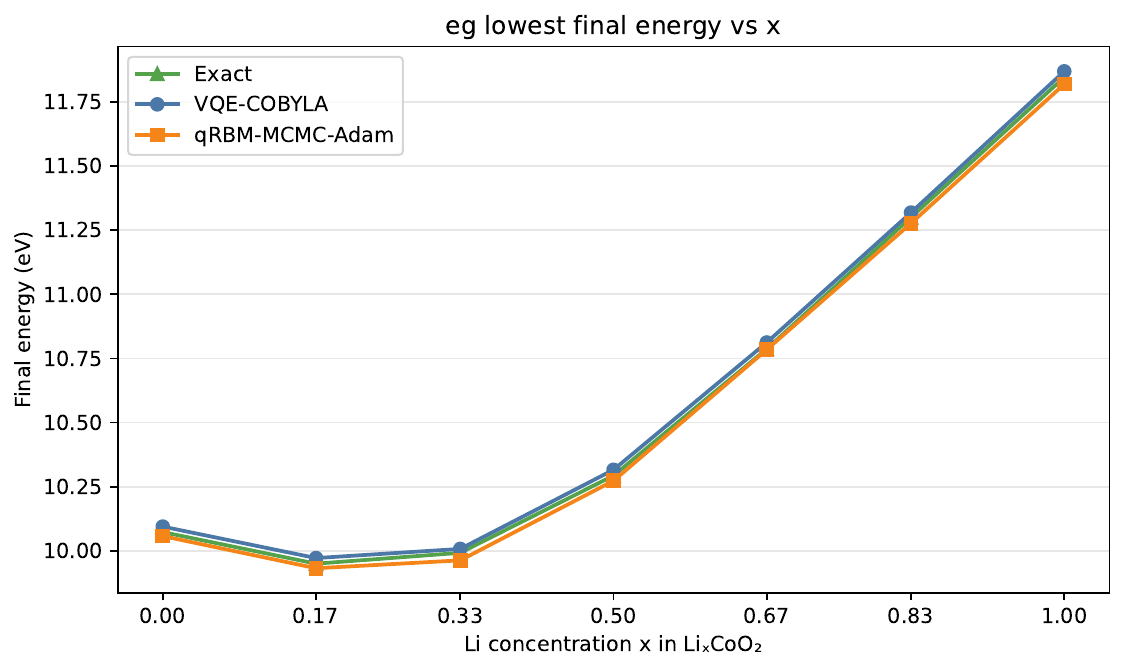}
        \caption{Lowest $e_g$ band-edge energy.}
    \end{subfigure}
    \hfill
    \begin{subfigure}[t]{0.49\linewidth}
        \centering
        \includegraphics[width=\linewidth]
        {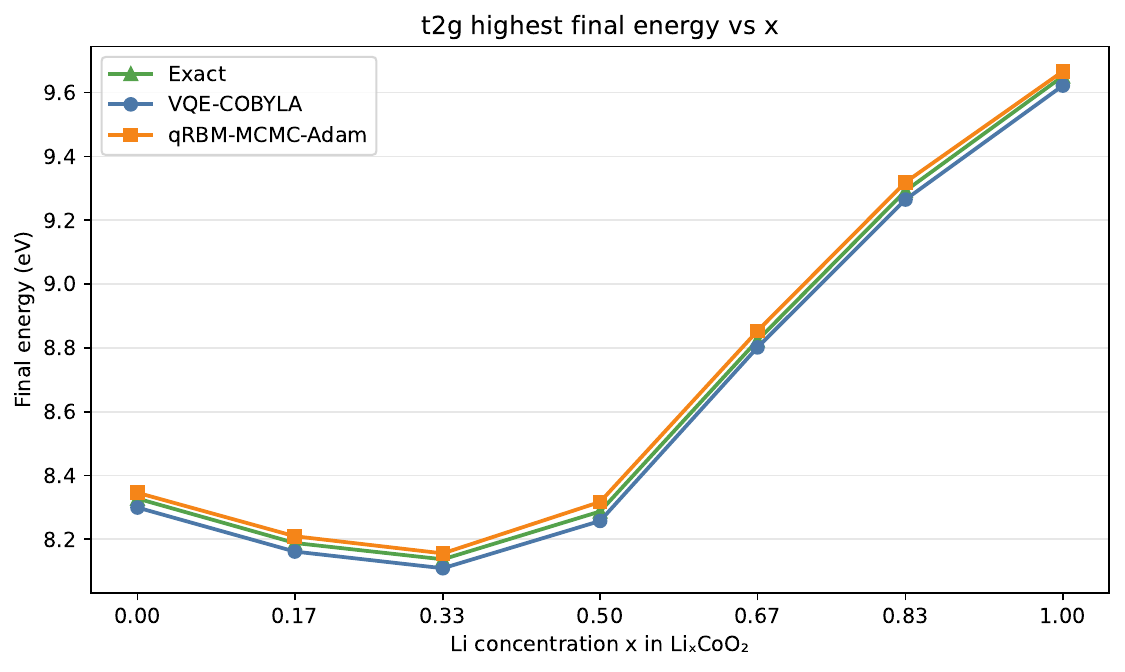}
        \caption{Highest $t_{2g}$ band-edge energy.}
    \end{subfigure}
    \caption{
    Composition-dependent evolution of the band-edge energies in
    Li$_x$CoO$_2$. Exact embedded-Hamiltonian eigenvalues are compared with
    HEA/VQE/COBYLA and NQS/VMC/Adam results for (a) the lowest $e_g$ state and
    (b) the highest $t_{2g}$ state. The distinct evolution of the two
    manifolds with lithium concentration illustrates the non-rigid
    electronic-structure response to delithiation.
    }
    \label{fig:si_composition_edges}
\end{figure}

To verify that the Hamiltonian family used in the transfer studies retains
the relevant electronic-structure behavior of Li$_x$CoO$_2$, we examined
the composition dependence of the highest $t_{2g}$ and lowest $e_g$
band-edge energies across the seven lithium concentrations considered in
this work. As shown in Fig.~\ref{fig:si_composition_edges}, both manifolds
evolve substantially with lithium concentration, but with distinct
composition-dependent shifts. This behavior is inconsistent with a simple
rigid-band picture and is consistent with the non-rigid electronic-structure
evolution reported during delithiation
~\cite{ensling2014nonrigid,tong2025first,xiong2012atomic}.

The HEA/VQE/COBYLA and NQS/VMC/Adam results closely follow the corresponding exact
solutions across the full composition range, demonstrating that the
variational calculations reproduce these composition-dependent electronic
trends. For fully lithiated LiCoO$_2$ ($x=1.00$), the embedded Hamiltonians
yield direct and indirect band gaps of 2.24 and 2.19~eV, respectively,
within the range of experimentally reported values
~\cite{kushida2001optical,liu2015electronic,ghosh2007structure,
balakrishnan2019studies,rao2009optical,van1991electronic}.
These results establish the physical relevance of the Hamiltonians used
for the transfer studies discussed in the main text.

\section{Supplementary Transfer and Accuracy Results}

\subsection{Band Energies Without Parameter Transfer}

\begin{figure}[htbp]
    \centering
    \includegraphics[width=\linewidth]
    {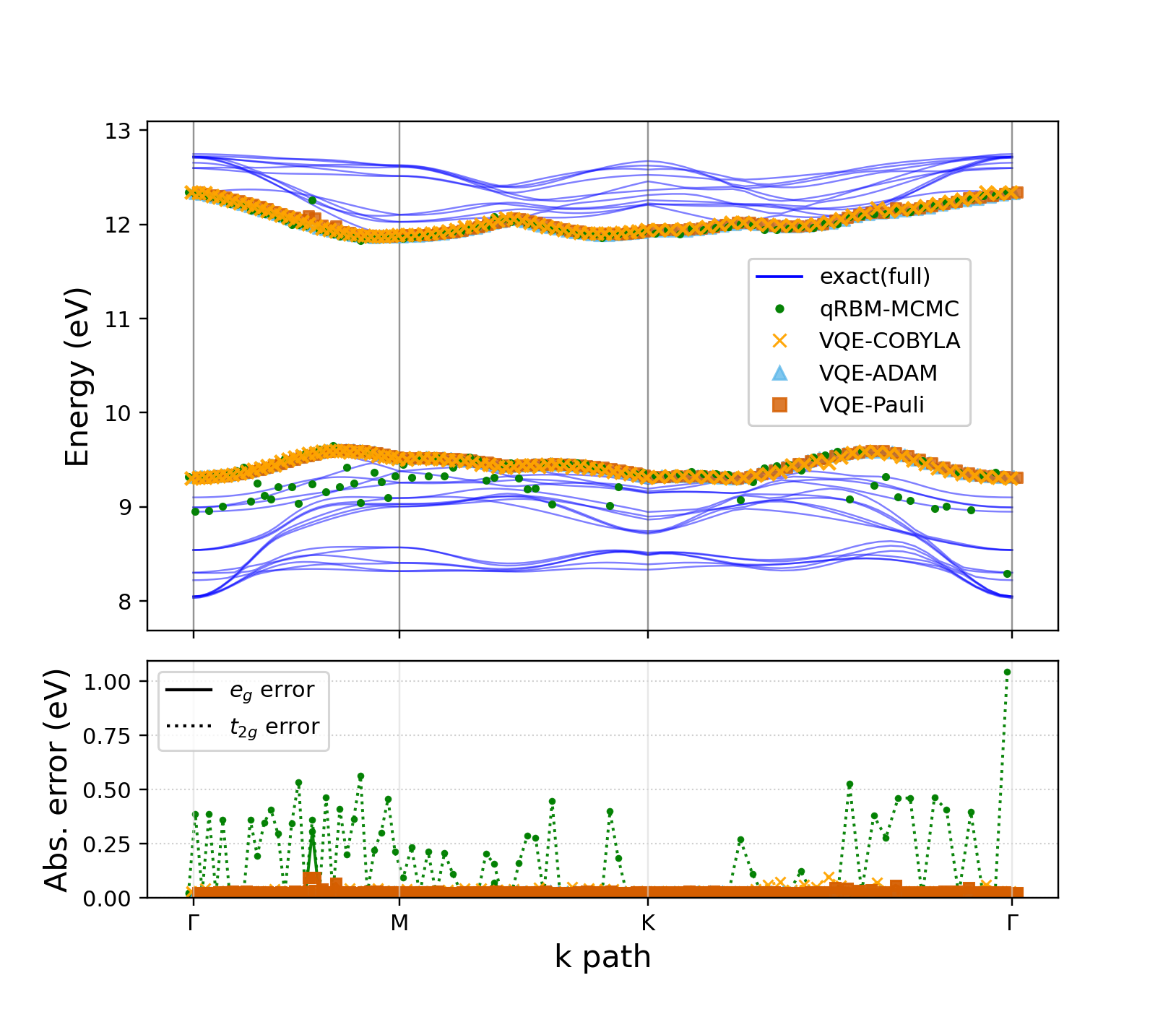}
    \caption{
    Band-energy comparison for four variational solver workflows using
    independent cold-start initialization.
    Exact full-band results obtained by diagonalizing the embedded
    Hamiltonians are shown as blue lines. The highest $t_{2g}$ and lowest
    $e_g$ band-edge energies obtained using NQS/VMC/Adam, HEA/VQE/COBYLA,
    HEA/VQE/Adam, and HPA/VQE/Adam are overlaid as markers.
    The lower panel reports the corresponding absolute band-edge errors.
    Results are shown along the 91-point
    $\Gamma$--$M$--$K$--$\Gamma$ path for
    fully lithiated LiCoO$_2$ ($x=1.00$).
    }
    \label{fig:si_all_methods_no_transfer}
\end{figure}

\clearpage




\subsection{Terminal Error Distributions}

\begin{figure}[htbp]
    \centering
    \includegraphics[width=\linewidth]
    {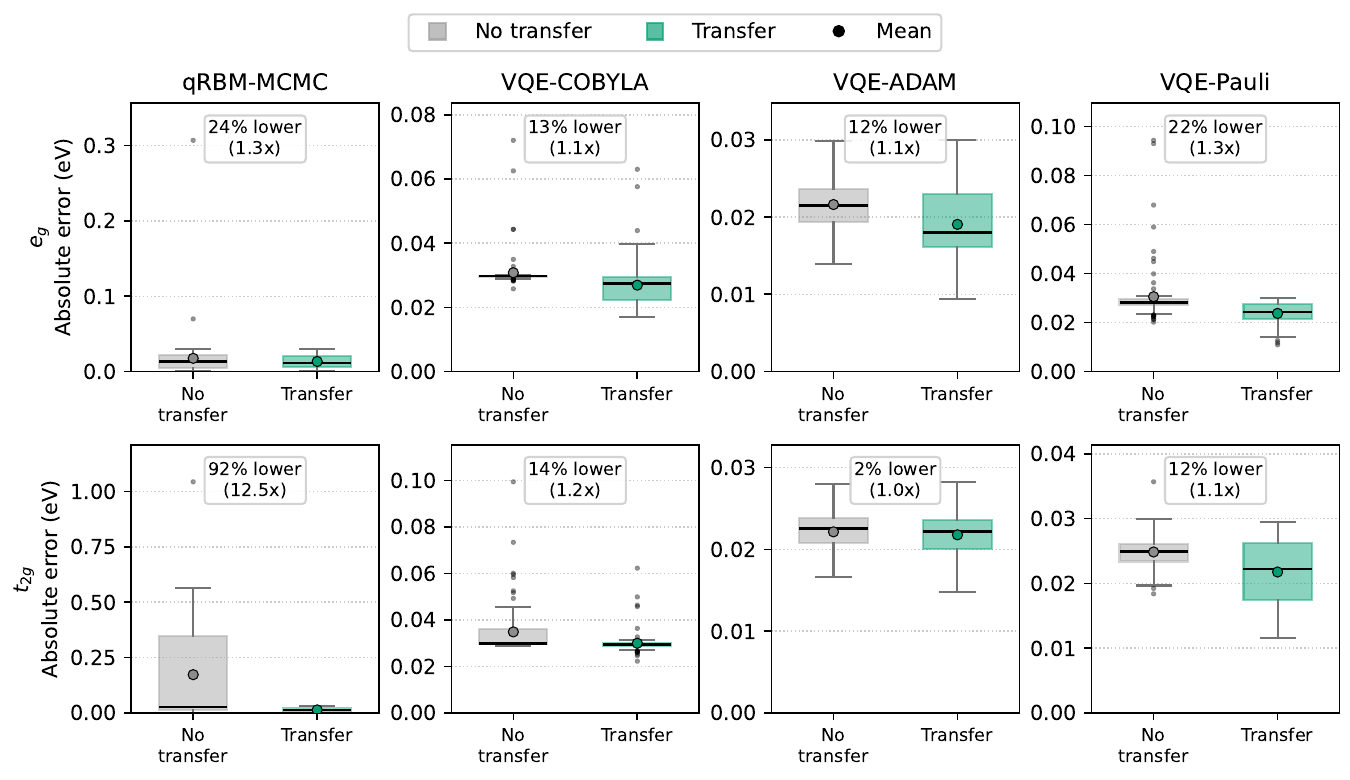}
    \caption{
    Distributions of absolute band-edge errors obtained with and without
    parameter-transfer initialization.
    Columns correspond to NQS/VMC/Adam, HEA/VQE/COBYLA, HEA/VQE/Adam, and HPA/VQE/Adam,
    while the upper and lower rows correspond to the $e_g$ and $t_{2g}$
    manifolds, respectively.
    Black circles denote mean errors. The annotated percentages and ratios
    report the reduction in mean absolute error obtained using transfer
    relative to the corresponding no-transfer calculation.
    }
    \label{fig:si_error_transfer_comparison}
\end{figure}